\documentclass[10pt]{article}

\pdfoutput=1

\usepackage{a4wide}
\usepackage{graphicx}

\usepackage{amssymb,amsfonts,amsmath}
\usepackage{marvosym}
\usepackage{wasysym}

\usepackage{hyperref}

\usepackage{tabularx}
\newcolumntype{L}[1]{>{\raggedright\arraybackslash}p{#1}}
\newcolumntype{C}[1]{>{\centering\arraybackslash}p{#1}}
\newcolumntype{R}[1]{>{\raggedleft\arraybackslash}p{#1}}

\let\oldepsilon\epsilon
\let\epsilon\varepsilon
\let\varepsilon\oldepsilon

\begin{document}

\begin{center}   
\textbf{\LARGE A lower-than-expected saltation threshold at Martian pressure and below}

\vspace*{0.2cm}

B. \textsc{Andreotti}$^a$, P. \textsc{Claudin}$^b$, J.J. \textsc{Iversen}$^c$, J.P. \textsc{Merrison}$^c$ and K.R. \textsc{Rasmussen}$^{c,d}$
\end{center}

{\small
\noindent
$^a$ {Laboratoire de Physique de l'Ecole Normale Sup\'erieure, UMR 8023 ENS -- CNRS -- Universit\'e de Paris -- PSL Research University, 75005 Paris, France.}\\
$^b$ {Physique et M\'ecanique des Milieux H\'et\'erog\`enes, UMR 7636 CNRS -- ESPCI Paris -- PSL Research University -- Sorbonne Universit\'e -- Universit\'e de Paris, 75005 Paris, France.}\\
$^c$ {Department of Physics and Astronomy, Aarhus University, 8000 Aarhus C, Denmark.}\\
$^d$ {Department of Geoscience, Aarhus University, 8000 Aarhus C, Denmark.}
}

\begin{abstract}
Aeolian sediment transport is observed to occur on Mars as well as other extraterrestrial environments, generating ripples and dunes as on Earth. The search for terrestrial analogues of planetary bedforms, as well as environmental simulation experiments able to reproduce their formation in planetary conditions, are powerful ways to question our understanding of geomorphological processes towards unusual environmental conditions. Here, we perform sediment transport laboratory experiments in a closed-circuit wind tunnel placed in a vacuum chamber and operated at extremely low pressures to show that Martian conditions belong to a previously unexplored saltation regime. The threshold wind speed required to initiate saltation is only quantitatively predicted by state-of-the art models up to a density ratio between grain and air of $4 \times 10^5$, but unexpectedly falls to much lower values for higher density ratios. In contrast, impact ripples, whose emergence is continuously observed on the granular bed over the whole pressure range investigated, display a characteristic wavelength and propagation velocity essentially independent of pressure. A comparison of these findings with existing models suggests that sediment transport at low Reynolds number but high grain-to-fluid density ratio may be dominated by collective effects associated with grain inertia in the granular collisional layer.
\end{abstract}

\vspace*{0.5cm}
\begin{center}
Proc. Natl. Acad. Sci. USA \textbf{118}, e2012386118 (2021).\\
\href{https://doi.org/10.1073/pnas.2012386118}{\texttt{https://doi.org/10.1073/pnas.2012386118}}
\end{center}
\vspace*{0.5cm}

%_____________________________________________________________________________
{W}ell-resolved satellite images and local pictures taken by rovers have provided multiple observational evidence of aeolian sand transport on Mars \cite{sullivan2008wind,hansen2011seasonal,bridges2012planet,lapotre2018curiosity,baker2018coarse,baker2018bagnold}. The surface of the planet is continually reshaped by wind-induced sand transport, forming ripples and dunes at decimetre, metre and hectometre scales \cite{lorenz2014elevation,lapotre2016large,silvestro2016dune,lapotre2018morphologic,vinent2019unified}. The sediment fluxes, estimated from the detected motion of ripples and dunes \cite{bridges2012earth,ayoub2014threshold,baker2018bagnold}, are only ten times smaller than in terrestrial deserts -- typically $2$--$20$~m$^2$/year. This observation seems at odds with the low CO$_2$ atmospheric pressure $P$ at the Martian surface, which varies from $6$ to $10$~hPa, depending on the season \cite{newman2017winds}. It is surprising that Martian winds can be strong enough \cite{iversen1982saltation,kok2010difference,berzi2017threshold} to set rather dense grains (basaltic material of typical density $\rho_p \approx 3 \times 10^{3}$~kg/m$^3$) in motion, even though gravity is almost three times smaller than on Earth ($g=3.7$~m/s$^2$). The description of aeolian saltation in such low-pressure conditions has remained controversial: conflicting theories either predict anomalously giant \cite{parteli2007saltation,almeida2008giant,sullivan2017aeolian,sullivan2020broad} or Earth-like centimetre scale trajectories \cite{ungar1987steady,andreotti2004two}. To shed light on this controversy \cite{lorenz2020martian}, here, we directly investigate sediment transport threshold \cite{greeley1976mars,greeley1980threshold,swann2020experimentally} in a controlled wind tunnel experiment \cite{creyssels2009saltating,ho2011scaling,li2012boundary}, varying the pressure \cite{merrison2008environmental} over three orders of magnitude. We also quantitatively compare the characteristics of impact ripples \cite{andreotti2006aeolian} with state of the art modeling of ripple formation and migration.

%------------------------------------------------
\section{Dimensionless numbers for sediment transport}
The comparison of sediment transport in terrestrial and Martian environments, and the experimental reproduction on Earth of  Mars-like conditions, require the identification of the relevant dimensionless numbers for sediment transport. The shear velocity $u_*$ can be rescaled using the grain diameter $d$ and the fluid mass density $\rho_f$ to form the Shields number
\begin{equation}
\Theta =\frac{ \rho_f u_*^{2}}{ \left( \rho_p- \rho_f \right) gd},
\end{equation}
which compares the fluid shear stress exerted on the particles at the surface of the bed to their apparent weight. As cohesion between grains is usually negligible for sand, the threshold Shields number $\Theta_t$ associated with the incipient motion of particles at the surface of a static bed only depends on a single dimensionless parameter: the Galileo number
\begin{equation}
{\mathcal G}=\frac{\sqrt{ \rho_f \left( \rho_p- \rho_f \right) gd^{3}}}{ \eta },
\end{equation}
which compares gravity to fluid viscosity $\eta$ effects. ${\mathcal G}$ can be interpreted as a Reynolds number based on the settling velocity; ${\mathcal G}^{2/3}$ can also be thought of as a dimensionless grain diameter.

This threshold has been extensively measured for incipient subaqueous bedload, varying grain diameter and fluid viscosity \cite{yalin1979inception,buffington1997systematic}. In that case, the average threshold data compares quantitatively with a simplified model based on a free body diagram on a single grain \cite{claudin2006scaling}. Because grain inertia is not included in this static force balance, the gravity term, proportional to $\left( \rho_p- \rho_f \right) gd^{3}$ is the only one involving particle density. A rigorous dimensional analysis implies that the static threshold curve $\Theta_t$ \textit{vs} ${\mathcal G}$ must be `universal' in the sense that it is valid for any environment.

However, the threshold shear velocity for aeolian transport, defined as the value below which already saltating grains colliding with the bed cannot sustain transport, is typically half that for subaqueous transport \cite{chepil1945dynamics,iversen1976saltation,iversen1994effect}. Its value must therefore depend on a second dimensionless parameter, the density ratio $\rho_p / \rho_f$, which can be varied continuously through dense (liquid) to dilute (gas) fluid conditions. Alternatively, any combination of ${\mathcal G}$ and $\rho_p / \rho_f$ can be used as a second controlled parameter, for instance the gravitational Stokes number $\mathcal{D}$ defined by:
\begin{equation}
\mathcal{D}=  \frac{\sqrt{\rho_p (\rho_p-\rho_f)gd^3}}{\eta}
\end{equation}
$\mathcal{D}$ gives the relative magnitude of gravity and viscous drag for a grain moving at the surface of the bed.

In the aeolian case, rather than a static force balance, a dynamic model, which idealizes saltation by means of a typical grain trajectory, is needed to reproduce those threshold data. Importantly, these static \emph{vs} dynamic designations for thresholds are theoretical concepts and are not used here in their pragmatic experimental acceptations (Supplementary Information (SI) Appendix). In our experiment, no hysteresis in sediment transport regime is observed upon increasing or decreasing the wind speed, so that a unique wind velocity threshold $u_t$ has unambiguously been measured using the procedure described below. As in air at ambient pressure, it must be compared to a dynamical threshold model.

\begin{figure}[p]
\centerline{\includegraphics{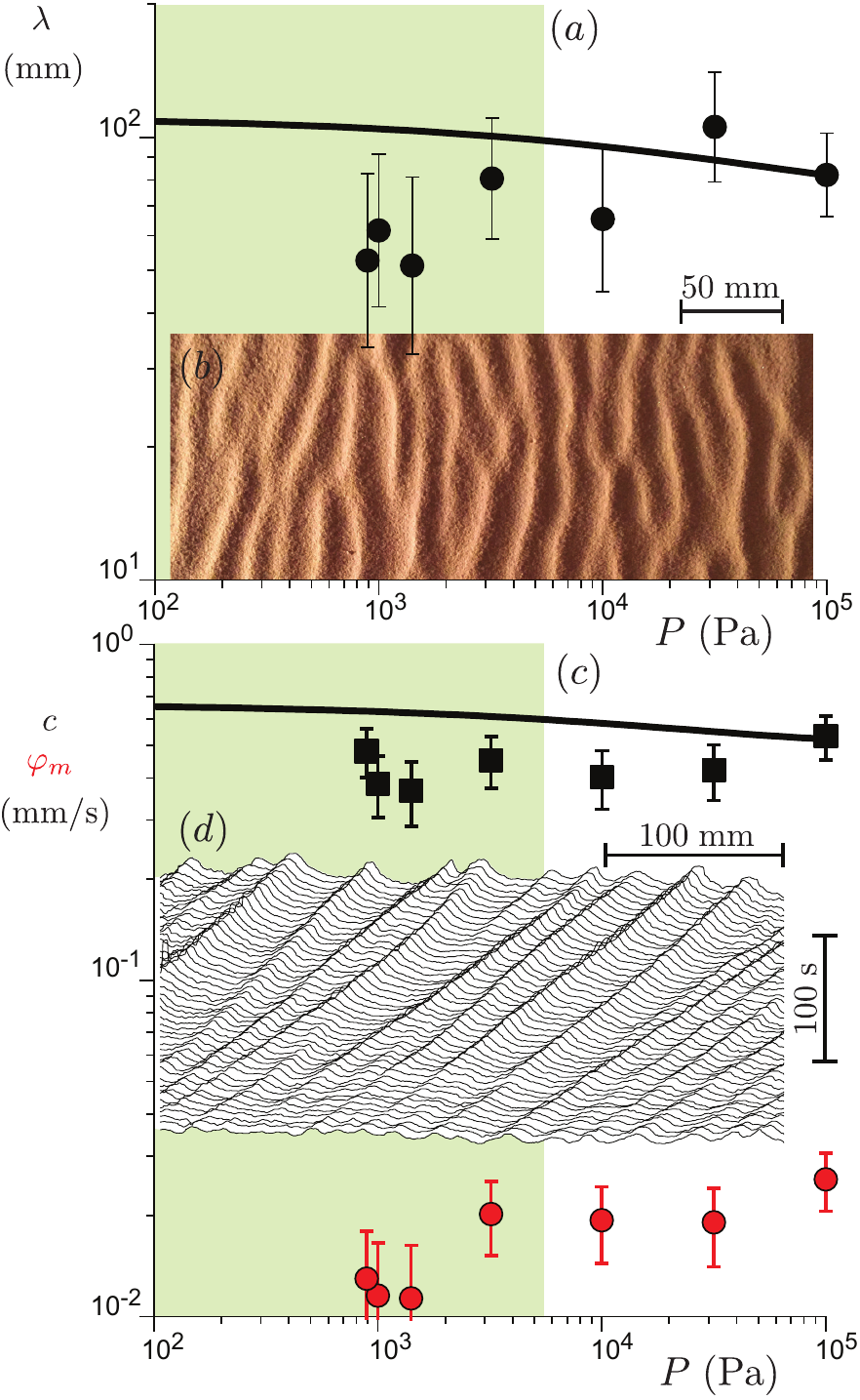}}
\caption{Impact ripple characteristics \textit{vs} pressure $P$. (a) Emergent wavelength $\lambda$ (black circles). (b) Photograph of the ripples in the tunnel at $u_*/u_t=1.1$. (c) Propagation speed $c$ (black squares) and maximum net bed erosion rate $\varphi_{m}$ (red circles). (d) Space-time diagram of the ripple elevation profiles showing their coarsening dynamics when the wind speed is suddenly increased from $u_*/u_t=1.1 $ to $1.5$. Time goes from bottom to top. Wind is from left to right.These are data in CO$_2$ Martian conditions. Solid lines in panels (a) and (c): adjustment of Eqs. (\ref{Eqc},\ref{Eqlambda}) with factors calibrated on independent data \cite{andreotti2006aeolian}. Statistical error bars corresponding to data dispersion on independent measurements are displayed.}
\label{Fig1}
\end{figure}
\begin{figure}[p]
\centerline{\includegraphics{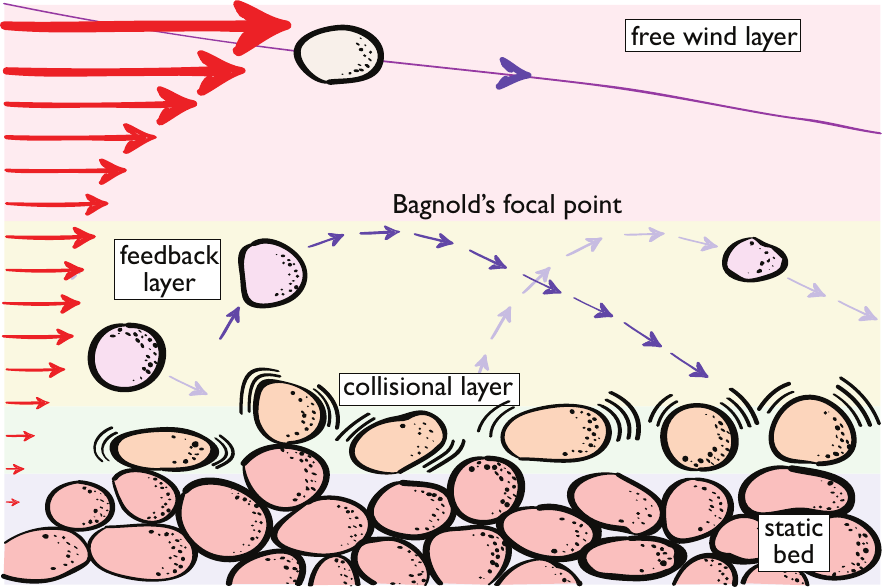}}
\caption{Three-layer picture of aeolian sediment transport. The main, central layer is the feedback layer, where the saltating grains slow down the wind, and their feedback ensures a unit replacement capacity during rebounds \cite{ungar1987steady}. Above Bagnold's focal point, defined as the top of that central layer, this feedback is negligible and the wind is unperturbed (`free wind' layer). At the interface with the static bed, momentum is transferred by collisions between the grains \cite{duran2014direct} (`collisional' layer). Note that, in this schematic, the vertical axis is not to scale: the altitude of Bagnold's focal point is on the order of $50d$, or a centimeter \cite{creyssels2009saltating,kok2012physics,pahtz2020physics}.}
\label{Fig2}
\end{figure}
%

%------------------------------------------------
\section{Low-pressure sediment transport experiments}
To investigate the role of $\rho_p / \rho_f$ in a systematic way, we performed controlled experiments in a pressure-controlled wind tunnel \cite{merrison2008environmental} (Methods, SI Appendix, Figs. \ref{fig:exp-setup-general}, \ref{fig:exp-setup-insert}). The working section of the tunnel is $0.85$~m high, which is larger than the highest trajectories. The width ($0.36$~m) is much larger than both the viscous and turbulent boundary layers, providing a homogeneous central region, where measurements are performed. The $3.25$~m length is comparable to that of working sections previously employed in low pressure saltation experiments. These finite sizes do not allow for fully developed trajectories as in an unbounded environment, but should not affect the value of the threshold, nor the characteristics of incipient ripples. At constant temperature, the fluid density $\rho_f$ is proportional to the pressure $P$. At low pressure, a larger wind velocity is required to transport the grains; this constraint determines in practice a minimum working pressure of around $1.7$~hPa for our experiments. The Martian density ratio $\rho_p / \rho_f \approx 1.6 \times 10^{5}$ has been reproduced both in air at $P \approx 14$~hPa and in CO$_2$ at $P \approx 9$~hPa. A granular bed made of quartz grains of diameter $d=125$~$\mu$m was used, equivalent to Galileo numbers of $1.9$ in CO$_2$ and $1.6$ in air and therefore to an equivalent grain diameter for Mars of $160$ and $140$~$\mu$m, respectively (SI Appendix, Supp. Tab. \ref{AmbientMarsDataTable}). During a typical experiment, the wind was blown for one minute at ambient pressure to smooth the flat sand bed before pumping the atmosphere to decrease the pressure to the desired value. A centimetre-thick saltation layer and emergent ripples were directly observed over three orders of magnitude in $P$.

%------------------------------------------------
\section{Impact ripple characteristics}
In Martian conditions, ripples forming just above the wind velocity threshold $u_t$ presented the same morphology as their terrestrial counterpart. They can unambiguously be identified as impact ripples, as one can continuously observe the formation of the same pattern when gradually decreasing the pressure. The space-time diagram of Fig. 1d shows the dynamics of the ripples when the wind speed is increased: the pattern propagates downwind and coarsens with a growing wavelength. We could not observe the final, saturated state of the ripples \cite{andreotti2006aeolian}, and focus on the emergent pattern characteristics. Importantly, Fig. 1 shows that neither the initial wavelength $\lambda$ of these incipient ripples nor their propagation velocity $c$ strongly depend on the pressure, at fixed $u_* / u_t$.

This lack of dependence on $\rho_p / \rho_f$ provides a strong test of the current understanding of aeolian sediment transport \cite{duran2011aeolian,kok2012physics,valance2015physics} (see also SI Appendix). The key idea is the existence of a saltation layer inside which the wind speed profile is reduced to its threshold due to the negative feedback of the particles on flow momentum (`feedback' layer, Fig. 2) \cite{bagnold1941book,owen1964saltation,mcewan1993kink,ungar1987steady,andreotti2004two,duran2011aeolian,valance2015physics}. The number of particles transported increases with wind speed but the characteristics of grain trajectories remain those of threshold conditions. However, this layer alone cannot explain the observed increase of ripple wavelength with $u_*$ . Numerical simulations \cite{duran2014direct} have revealed the existence, at the interface with the static bed, of a collisional layer (Fig. 2) responsible for this behavior. Contrary to the original idea of saltating grains splashing and ejecting grains from the static bed, saltation transmits its momentum to a whole quasi-two-dimensional layer governed by gaseous-like collective effects. The grain velocities in this layer therefore behave differently than in the feedback layer and depend on $u_*$ (SI Appendix). The mathematical modelling of the coupling between the feedback and collisional layers provides the following scaling laws for the ripple velocity and wavelength (see \cite{duran2014direct} and Methods):
\begin{eqnarray}
 c &\sim& u_t\sqrt{\frac{ \rho_f}{ \rho_p}}\sqrt{\frac{u_*^{2}}{u_t^{2}}-1}, \label{Eqc}\\
  \lambda &\sim& u_t\sqrt{\frac{d}{g}}\sqrt{\frac{ \rho_f}{ \rho_p}}\sqrt{\frac{u_*^{2}}{u_t^{2}}-1}. \label{Eqlambda}
\end{eqnarray}
These relationships reproduce well the increase of $c$ and $\lambda$ with the wind speed in a quasi-linear fashion \cite{andreotti2006aeolian}. Their proportionality factors, which only depend on the grain characteristics but not on the fluid density, are calibrated using ambient pressure data (SI Appendix, Supp. Fig. \ref{fig:RippleScaling}). The predictions of Eqs. (\ref{Eqc},\ref{Eqlambda}) are superimposed on the data in Fig. 1a,c and show fair agreement. As the threshold shear stress $\rho_f u_t^{2}$ is, to a first approximation, independent of $P$ (Fig. 3d), Eqs. (\ref{Eqc},\ref{Eqlambda}) indeed predict that $c$ and $\lambda$ remain roughly constant at fixed $u_* / u_t$, as evidenced over the three orders of magnitude of pressure investigated here. This is a key result to interpret the smallest decimeter-scale Martian bedforms as impact ripples. A closer look to the data shows a slight increasing trend of the wavelength with pressure, which would need confirmation with additional experiments to reduce error bars on measurements.

Close to the threshold, the air flow entering the working section of the tunnel is free of particles. As grains are entrained into motion, the sediment bed is eroded and the sediment flux progressively increases downwind until it reaches saturation. At ambient pressure, the erosional zone is limited to the entrance bed region, corresponding to a metre-scale saturation length \cite{andreotti2010measurements}. Upon decreasing $P$ , this zone extends downwind, and below 300~hPa, it encompasses the entire bed. The maximum net erosion rate $\varphi_{m} $ along the surface remains much smaller than grain velocity in the collisional layer (on the order of $c$), and, like ripple characteristics, does not depend much on $P$ (Fig. 1). Considering that both feedback and collisional layers rapidly equilibrate to the wind conditions independently of $P$, the overall erosion of the bed remains to be explained. The feedback layer with grains experiencing a flow at the threshold is unstable \cite{andreotti2004two}: any grain bouncing above Bagnold's focal point experiences a larger drag and is accelerated to velocities comparable to that of the wind. The bed erosion is therefore associated with the gradual mobilisation of grains towards the upper part of the transport layer, or `free wind' layer (Fig. 2). In the case of saturated transport, this upper layer is in dynamic equilibrium with the lower feedback layer through a balanced net flux. The saturation of the free wind layer is controlled by the drag length, defined as the distance needed to accelerate a grain to the velocity of the undisturbed wind, which scales as $\frac{\rho_p}{ \rho_f} d$ in the inertial regime \cite{andreotti2010measurements}. The overall erosion of the bed in the wind tunnel under Martian conditions affects the morphology of the impact ripples, which display sharp slightly segregated crests, where the coarser grains accumulate.

\begin{figure*}[p]
\centerline{\includegraphics[width=\linewidth]{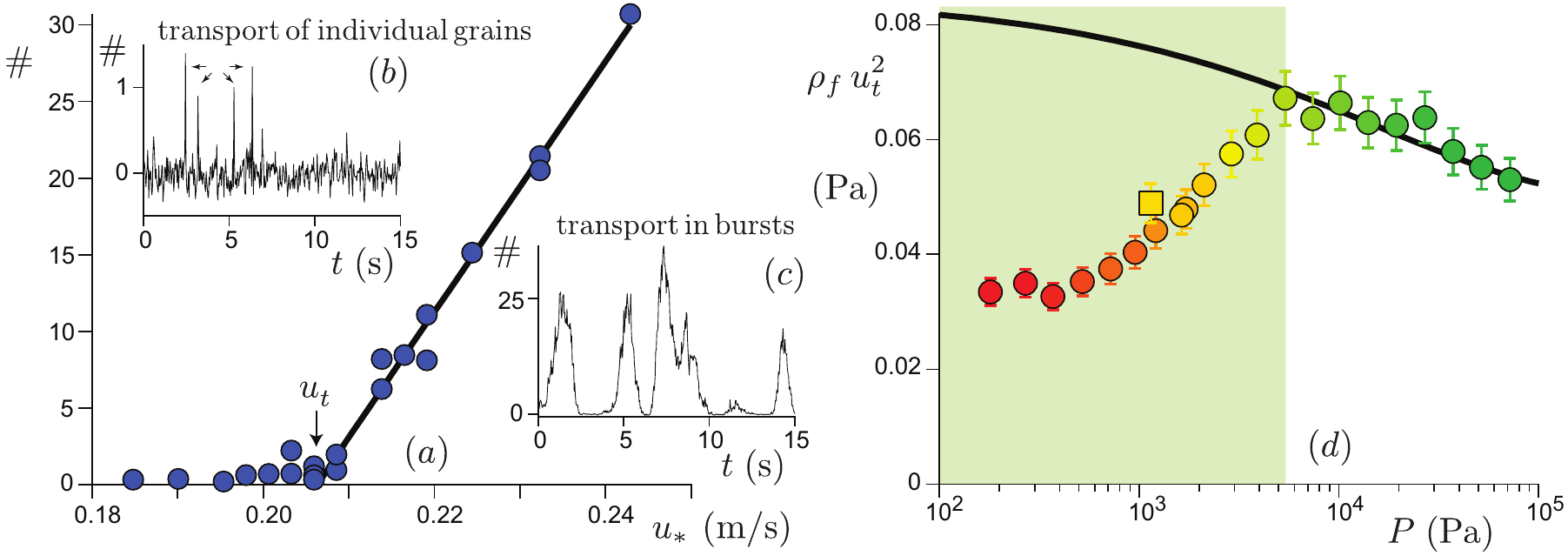}}
\caption{Sediment transport threshold. (a) Averaged number of grains passing through the control window during a $30$~s time interval \textit{vs} wind speed. The threshold $u_t$ is the cross-over between a regime of intermittent individual grains (arrows in (b)) and a fluctuating but steady transport (c). The precision on the number of grains is of the size of symbols; dispersion of data indicates the repeatability. (d) Threshold shear stress $\rho_f u_t^{2}$ \textit{vs} pressure $P$ (filled circles). Square: experiment in $CO_2$, in Martian conditions. All these data are for quartz grains of size $d=125$~$\mu$m. Solid line: prediction of the model (Methods, SI Appendix) adjusted on subaqueous and aeolian data, for different $d$ at ambient pressure. Green background: zone corresponding to the new low-pressure regime (large density ratio $\rho_p / \rho_f$). The $7$ \% error bars reflect both measurement statistical errors on the threshold and uncertainties on the relation between shear velocity and rotation speed.}
\label{Fig3}
\end{figure*}
%

%------------------------------------------------
\section{Saltation threshold}
Although the fact that sand transport occurs on Mars is now well established, the value of the threshold wind velocity $u_t$ as a function of grain and fluid characteristics has remained controversial. Here, we use a long-distance microscope to measure the number of moving grains in a control volume located close to the bed as a function of time (Methods, SI Appendix). These measurements are performed on a levelled granular surface, without ripples, outside the lateral boundary layers. With increasing wind speed, one observes a transition from a regime of individual grains transported intermittently due to turbulent fluctuations to a regime with bursts composed of a number of moving particles that is rapidly increasing with the shear velocity (Fig. 3a, SI Appendix, Supp. Fig. \ref{fig:exp-videogram}). This behaviour is typical of an imperfect bifurcation in the presence of noise, for which the threshold can be defined by extrapolating to zero the average number of grains per burst. Below this threshold, the overall sediment flux decreases rapidly and is negligible. Therefore, its feedback on the wind velocity profile can be neglected. Calibrations of the shear velocity on a rigid bed can then be used safely (SI Appendix). This analysis provides accurate and reproducible values of $u_t$. The quantification of the transition between transport regimes allows us to measure a threshold with error bars significantly smaller than those reported in previous experiments \cite{greeley1980threshold,swann2020experimentally} (see SI Appendix, Supp. Fig. \ref{fig:Swann} for a comparison).

The corresponding behaviour of the threshold wind shear stress $\rho_fu_t^{2}$ with respect to $P$ is shown in Fig. 3d. At relatively high pressures ($P \gtrsim 60$~hPa), results follow the prediction of a `dynamic' threshold model adjusted in the ambient conditions \cite{andreotti2004two,claudin2006scaling,duran2011aeolian,claudin2017dissolution,vinent2019unified}. It accounts both for the transition from Stokes to turbulent drag as determined by the grain Reynolds number, and for the transition from a rough to a smooth (viscous) boundary layer (Methods, SI Appendix). At low pressures, however, the measured threshold gradually deviates from this expected theoretical law, providing experimental evidence of a new regime, which cannot be explained by these hydrodynamic transitions. Thanks to the unprecedented accuracy ($7$\%, see SI Appendix for calibrations) the drop by a factor of $\simeq 2$ in threshold stress in this regime is unambiguously resolved in our experiment. This change of regime can also clearly be observed on raw data (SI Appendix, Supp. Fig. \ref{fig:RawData}), which confirms that it does not result from a calibration artefact. This behaviour does not coincide with the Knudsen regime: the molecular mean free path remains, in the investigated pressure range, much smaller than the grain size and than the viscous sub-layer \cite{jakobsen2019laboratory}. We have also excluded other possible spurious effects like electrostatics or finite size effects, which could putatively lower the transport threshold.

In order to isolate the effect of the density ratio, the threshold must be represented in the dimensionless plane relating the Shields number to the Galileo number (Fig. 4). Compiled measurements of the threshold for subaqueous bedload \cite{yalin1979inception,gaucher2010experimental} robustly follow the theoretical static curve (Methods). Lowering ${\mathcal G}$ from large values, one may observe several well known transitions: the decrease of the threshold around ${\mathcal G} \approx 10^{3} $ is associated with the change from turbulent to Stokes drag on the grains; its increase below ${\mathcal G} \approx 10^{2} $ is due to the transition from rough to smooth bed conditions. Cohesion raises the threshold at very small grain sizes (Supp. Fig. \ref{fig:Water}). Note that in our experiments, with grains of fixed diameter $d = 125 \mu$m, cohesion can be neglected. As it stays below the static threshold, aeolian data \cite{chepil1945dynamics,rasmussen1996saltation} are fitted by the model dynamic threshold. The offset between theoretical static and dynamic thresholds shows the importance of grain inertia to sustain saltation at wind speeds lower than the static threshold. In this closed-loop tunnel, we have not observed any significant transport hysteresis. The bed heterogeneities and the wind turbulent fluctuations are sufficient to eventually activate transport below the static model curve. This lack of hysteresis incidentally allows us to define unambiguously and accurately a unique transport threshold.

When varying the Galileo number by varying $\rho_p/\rho_f$ for a fixed grain size (here $d = 125 \mu$m), the dynamic model predicts a threshold that naturally coincides with the `universal' static curve at low $\rho_p / \rho_f$ (dense fluid) and progressively deviates from it to cross the ambient-aeolian curve at the point corresponding to that grain size. It further follows the trend of our low-pressure points, but only down to a value of $\mathcal{G} \approx 3$, below which a new regime takes place, where measured thresholds are significantly below the expected values. Previous experimental data obtained in the NASA Martian wind tunnel using walnut shells \cite{greeley1980threshold} of size $212$~$\mu$m in CO$_2$ are reasonably consistent with our measurements for the larger values of $\rho_p / \rho_f$, but crucially do not enter this low fluid density region. Lower-than-expected values of the threshold have also been found by \cite{swann2020experimentally} (SI Appendix, Supp. Fig. \ref{fig:Swann}), but their more qualitative detection as well as their data scatter over a factor of $\simeq 3$ make any quantitative comparison difficult. Our CO$_2$ data point in Martian conditions also collapses on the same curve in this dimensionless graph, in contrast with its dimensional representation as a function of pressure (Fig. 3).

\begin{figure}[p]
\centerline{\includegraphics{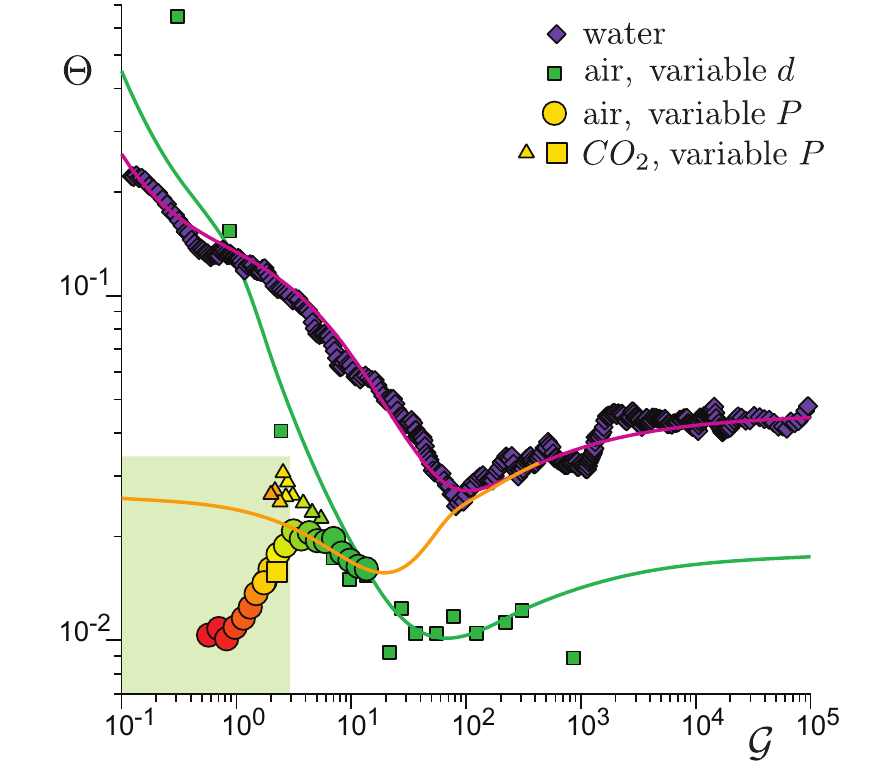}}
\caption{Transport threshold in the dimensionless plane Shields $\Theta$ \textit{vs} Galileo ${\mathcal G}$ numbers. Gathering of subaqueous (violet diamonds) and aeolian (green square) data of the literature \cite{chepil1945dynamics,yalin1979inception,rasmussen1996saltation,gaucher2010experimental} for sand grains with variable $d$ in ambient conditions, and our aeolian data (circles) with variable $P$ and fixed grain size $d=125$~$\mu$m. The symbols' colour codes for the density ratio $\rho_p / \rho_f$: from $1$ (violet) to $10^6$ (red), in log scale -- green is in the range $10^3$--$10^4$. Previously published data from the NASA wind tunnel using walnut shells \cite{greeley1980threshold} of size $212$~$\mu$m in CO$_2$ are displayed (triangles) too, see also Supp. Fig. \ref{fig:Greeley} for data corresponding to other grain diameters used in that paper. Solid lines: threshold predictions for subaqueous (purple), aeolian at ambient (green) and variable (orange) pressure conditions. Green background: new regime range.}
\label{Fig4}
\end{figure}
%

%------------------------------------------------
\section{Discussion}
The central result of this paper, beyond ripples properties being invariant with pressure, is the continuous transport threshold curve obtained with an unprecedented accuracy as a function of the Galileo number. It shows a regime at small $\mathcal{G}$ whose understanding is a challenge that must be tackled by further experiments, varying grain size and gas properties. From this perspective, it is instructive to exploite threshold values reported in \cite{greeley1980threshold} for the four smallest grain sizes at $\mathcal{G} \simeq 1$  (SI Appendix, Supp. Fig. \ref{fig:Greeley}). They suggest a threshold Shields number scaling as $\Theta \propto \mathcal{D}^{-1}$ in the low Galileo regime (SI Appendix, Supp. Fig. \ref{fig:Greeley_Geq1}). This condition is equivalent to a surface velocity gradient proportional to the typical granular collision frequency $\sqrt{g/d}$.

Novel ideas are required to explain the physical processes at work in this saltation regime reached at Galileo numbers orders of magnitude below well-known hydrodynamic transitions. It is also characterised by a reduced noise in grain trajectories and bursts (SI Appendix, Supp. Fig. \ref{fig:exp-burst-lowpressure}). We hypothesize that the physics in this regime is associated with mechanisms at the granular bed level. Two surface processes are currently not accounted for in the dynamic model to which data are compared. First, the surface is irregular at the grain scale and thus must affect grain rebound and trapping. Second, it ignores any collective effects, and in particular the collisional layer mentioned above, whose crucial role was previously emphasized for the ripple dynamics. One could imagine that a collisional layer, presenting a wide distribution of grain hop lengths \cite{duran2014direct}, is already needed above the threshold to sustain transport. Such a gaseous-like layer could play the role of granular temperature to activate grain motion.

Finally, the existence of this regime is of practical importance as it arises for large values of $\rho_p / \rho_f$ that are commonly encountered on planetary bodies with relatively thin atmospheres. Of particular importance, we predict this regime to dictate saltation thresholds on Mars. Because saltating grains may impact landed assets and lift dust, there is a critical need to further investigate the mechanics behind sand transport under Martian conditions to mitigate risks during future robotic and human missions. Even larger density ratios occur on comets, where transient thermal winds have been shown to be responsible for the emergence of giant ripples \cite{jia2017giant}. In denser atmospheric conditions, such as those on Titan and Venus \cite{burr2015higher,berzi2016periodic,pahtz2020unification}, for which the fluid density is intermediate between water and air (typically $\rho_p / \rho_f \approx 10^{2}$), our measurements predict that this regime to be irrelevant, so that the extrapolation of the model calibrated in  ambient conditions remains valid and can be used for quantitative predictions of sediment transport threshold and characteristics of emergent bedforms.

\vspace*{0.3cm}
%____________________________________________________________________________
\noindent
\rule[0.1cm]{3cm}{1pt}

This work has been funded by Europlanet grant No 11376. Europlanet 2020 Research Infrastructure has received funding from the European Union's Horizon 2020 research and innovation program under grant agreement No 654208. We are grateful to R.C. Ewing for providing walnut shell samples for a new measurement of their bulk density.

\newpage

\noindent
\textbf{\Large MATERIALS AND METHODS}\ \hrulefill \\

We describe here the experimental set-up, the measurements, as well as the theoretical framework for the computation of the transport threshold. More details, references, technical points as well as additional figures and calibration curves can be found in the Supplementary Information (SI) document.

%------------------------------------------------
\section*{Wind tunnel}
The wind tunnel set-up consists of a turbine section with a bi-propeller system. The rotation of the propeller fans at a controlled angular frequency generated two return flows above and below the working section (SI Appendix, Supp. Fig. \ref{fig:exp-setup-general}). The sand bed was prepared in the 3.25 m long tunnel, which had a rectangular cross section about 0.36 m wide and 0.85 m high (SI Appendix, Supp. Fig. \ref{fig:exp-setup-insert}). The container could be hermetically closed and the interior pressure adjusted to the desired value. Pressure calibration of the two low-pressure (Pirani gauge Pfeiffer TPR280) and high-pressure (capacitance gauge Pfeiffer APR250) sensors were made by successive injections of a given amount of gas in a chamber of known volume (SI Appendix, Supp. Fig. \ref{fig:exp-PressureCalibration}). The absolute pressure is measured within $2\%$ except below $500\;{\rm Pa}$, where uncertainties level off at $10\;{\rm Pa}$.

%------------------------------------------------
\section*{Saltation threshold}
A digital microscope was installed at 30 cm from the end of the working section, looking across the bed and focused around 12 cm into the wind tunnel. Here, a horizontal laser sheet crossed the microscope field of view (at around 45$^{\circ}$) and close to the sand bed surface (around 1 cm). The laser sheet was thick enough (1 mm) for saltating sand grains to be tracked (SI Appendix, Supp. Fig. \ref{fig:exp-videogram}). Image analysis then allowed us to quantify the number of particles passing through the field of view. The threshold was defined as the transition between saltation of groups of particles (bursts) to intermittent saltation of single particles (at high pressure) or no transport (at low pressure), see Fig. 3 and SI Appendix, Supp. Fig. \ref{fig:exp-burst-lowpressure}. It is measured on a flat bed, in a situation where the negative feedback of transport on flow velocity is negligible. The threshold fan angular frequency $\Omega_t$ is measured within $2.5\%$. Due to uncertainty on the calibration, the threshold shear velocity, $u_t$ is measured within $3.5\%$.

%------------------------------------------------
\section*{Bed erosion rate and ripples}
Four overhead cameras were mounted equidistant between the inlet and outlet of the working section while four sheet lasers were placed outside the tunnel impinging the sand bed close to the center of the tunnel and forming a straight line on an initially flat bed (SI Appendix, Supp. Fig. \ref{fig:exp-setup-insert}). The inclination of each laser sheet was adjusted to be around 15$^{\circ}$ from the horizontal. Changes in bed elevation resulted in a sideways displacement of the laser line, which was recorded by the cameras. A four-step vertical calibration target was placed on the bed below each camera and imaged in order to verify the conversion between transverse displacement and bed elevation. This system allowed continuous monitoring of changes in bed elevation with a resolution on the order of 0.08 mm (one pixel). From bed elevation, the erosion rate profile as well as the emerging ripple characteristics (wavelength, propagation velocity) were measured as functions of time and pressure (Fig. 1). The average erosion rate has been measured at different locations, showing a statistical dispersion around $0.005$~m/s.

%------------------------------------------------
\section*{Wind shear velocity}
Wind velocity profiles (SI Appendix, Supp. Fig. \ref{fig:WindProfilesForShearVelocity}) were measured in the tunnel by means of arrays of Pitot tubes. Calibration was made with a 2-D Dantec Flow-Lite laser-Doppler anemometer. To avoid possible clogging of the Pitot tubes by saltation grains, these measurements were systematically performed over a fixed bed made of glued sand grains. We checked over the available runs that they are consistent with corresponding measurements over mobile beds (SI Appendix, Supp. Figs. \ref{fig:ExPressureRecord}, \ref{fig:LDAvsPitot}, \ref{fig:UFixedGranularBed}).

These profiles were analysed using a hydrodynamic model where the turbulent boundary layer flow is described by means of a first order Prandtl type closure. To account for both smooth and rough bed regimes, we adopted a van Driest-like mixing length, where the various constants were calibrated independently. In the unbounded case the fluid shear stress $\tau_{f}$ is the constant $\rho _{f} u_*^{2}$, defining the shear velocity $u_*$. A phenomenological stress profile of the form
\begin{equation}
\tau_{f}= \rho _{f}u_*^{2}\mathrm{\exp } \left[ - \left( \frac{z}{ \delta } \right) ^{2} \right],
\end{equation}
where $z$ is the vertical distance to the bed and $\delta$ the thickness of the turbulent boundary layer, provides a robust fit to the data. $\delta$ is found on the order of a few cm in the measurement zone, gently increasing when decreasing the pressure from the ambient to a few hPa. The prediction of the hydrodynamic model is fitted to the data by adjusting $u_*$ and $\delta$, which allows us to calibrate the relation between  the ratio $u_*/\Omega$ and the Reynolds number $\rho_f u_* d/\eta$ (SI Appendix, Supp. Fig. \ref{fig:expustarsrpm}). The systematic error on $u_*$ can be decomposed into two parts: (i) the miscalibration as a whole (all values, regardless the pressure, would be over or under estimated) is smaller than $1$\%; (ii) the slope of the relation may also produce systematic variations with pressure, at worst on the order of $2.5$\%.

%------------------------------------------------
\section*{Transport threshold model}
The theoretical framework used here for the computation of the transport static and dynamic thresholds gather elements from previously published studies \cite{andreotti2004two,claudin2006scaling,duran2011aeolian,claudin2017dissolution}.

The static shear velocity threshold $u_{ts}$ is computed from the force balance on a grain at rest at the surface of the bed that opposes the drag force of the flow and the effective friction of the bed. We take the drag force on a grain as a function of the difference between the grain velocity $\vec{v}$ and the fluid velocity $\vec{u}$ at the grain's location. The drag coefficient depends on the particle's Reynolds number based on that velocity difference $R_{u}= |\vec{u}-\vec{v} | d \rho _{f} / \eta$ and accounts for both inertial and viscous (Stokes) regimes. Cohesion between grains due to Van der Waals adhesive contact forces are taken into account. The various parameters involved in this static balance are adjusted to reproduce the threshold curve for subaqueous bedload (Fig. 4).

For the dynamic threshold, the equation for the grain motion is integrated. Upon colliding with the bed, the grain is assumed to rebound with a given ejection angle and a velocity ratio. The criterion for steady transport at the dynamic shear velocity threshold $u_{t}$ is that the kinetic energy of the take off particle is just enough to escape from the potential traps between its neighbours. Taking into account gravity and the force exerted by the wind, the take-off velocity takes the form
\begin{equation}
v_{ \uparrow }=a\sqrt{gd \left( 1-\frac{u_{t}^{2}}{u_{ts}^{2}} \right) }.
\end{equation}
where $a\simeq 11$ is a dimensionless number obtained by fitting the aeolian data at ambient pressure. As expected, $v_{\uparrow }$ vanishes at the static threshold $u_{t}=u_{ts}$ . The various parameters involved in this dynamic analysis are fitted to reproduce the threshold curve in the case of aeolian saltation at ambient pressure (Fig. 4).

\newpage

\noindent
\textbf{\Large SUPPLEMENTARY INFORMATION}\ \hrulefill \\

As an extension of the Materials and Methods, this file gathers all technical details on the experimental set-up and calibrations. It also summarises the theoretical framework from the literature for the sediment transport and ripple models.

%______________________________
\section*{Wind tunnel and instruments }

%_______________________
\subsection*{Environmental wind tunnel}
\begin{figure}[p]
\begin{center}
\includegraphics[width=\linewidth]{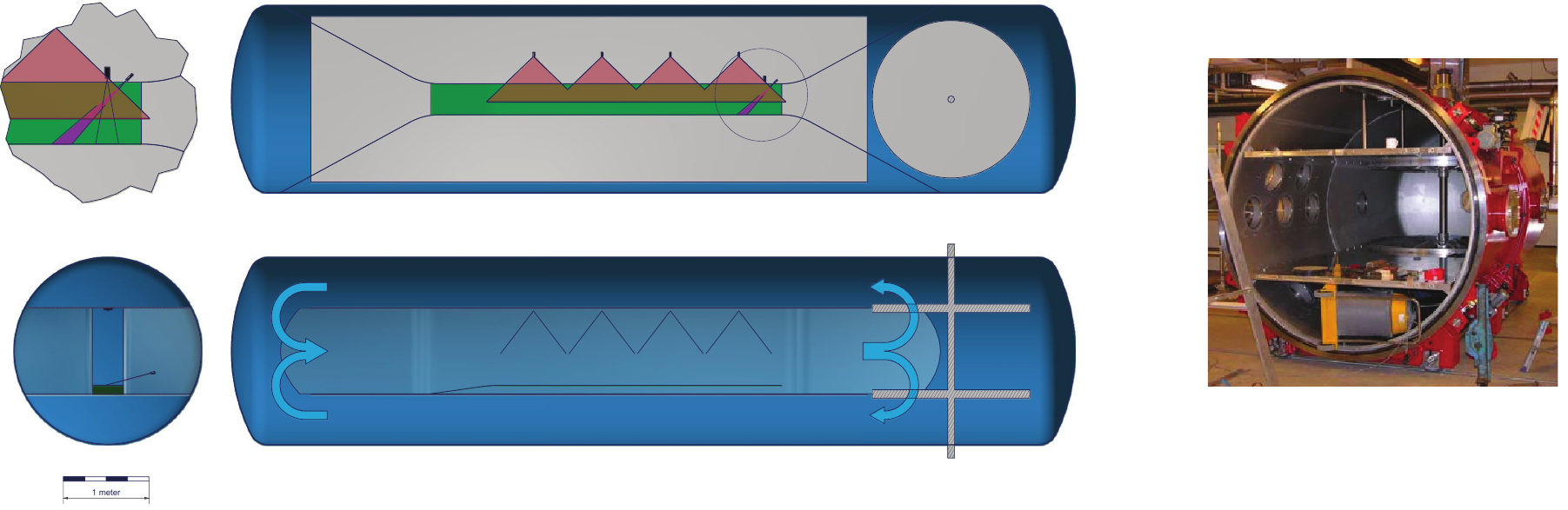}
\end{center}
\caption{Schematics of the set-up, with insert boundary layer tunnel. The working section is approximately $6$~m long, $2$~m wide and $1$~m high. Right: Photograph of the wind tunnel when open in halves.} 
\label{fig:exp-setup-general}
\end{figure}

As shown in Supp. Fig.~\ref{fig:exp-setup-general}, the tunnel set-up presents a working section which is $6$~m long, $2$~m wide and $1$~m high and a bi-propeller turbine section. The rotation of the propeller fans at a controlled frequency $\Omega$ (given in Fig.~\ref{fig:RawData} in rotations per minute (rpm)) generates a flow which is directed to the two return flow sections above and below the working section. At the upstream end of the working section, the flow from the two return sections is mixed in the entry section and directed into a contraction leading into the insert tunnel (Supp. Fig.~\ref{fig:exp-setup-insert}). The granular bed is prepared in the $3.25$~m long insert tunnel, inside the working section. The insert tunnel has a rectangular cross section around $0.36$~m wide and $0.85$~m high. The container can be hermetically closed and the interior pressure $P$ adjusted to the desired value.

%_______________________
\subsection*{Detecting saltation threshold}
A digital microscope is installed just upwind of the end of the working section looking across the bed and focused around $12$~cm into wind tunnel. Here, a horizontal laser sheet crossed the microscope field of view (at around $45^\circ$) and close to the granular bed surface (around $1$~cm). The laser sheet was thick enough ($\gtrsim 1$~mm) for saltating grains to be tracked.

At each working pressure, the fan frequency $\Omega$ was raised by steps from below threshold (no transport) to a value where visual inspection of the microscope images showed abundant saltation (Supp. Fig.~\ref{fig:exp-burst-lowpressure}c-e). $\Omega$ was then decreased in small steps until a few, if any, saltating particles were observed within a period of $15$-$30$~s. At each step in $\Omega$, the saltation intensity has been quantified from the video recording by estimating the number of particles passing through the microscope window (Fig.~2a Supp. Fig.~\ref{fig:exp-burst-lowpressure}). The threshold was defined as the transition between saltation of groups of particles (bursts) to intermittent saltation of single particles (at high pressure) or no transport (at low pressure). It must be emphasized that former studies have used other definitions of the transport threshold, leading to much larger dispersion of the data. Not only have the threshold measurements presented here been made quantitative, precise and repeatable, but also correspond to an objective change of regime detected by a visual criterion (Fig.~2a, Supp. Fig.~\ref{fig:exp-videogram}), thanks to the microscope imaging. Note also that, in a recirculating wind tunnel, there are always residual grains injected upstream of the working section. Threshold was measured for decreasing as well as increasing $\Omega$ corresponding to two estimates of the wind shear velocity $u_{t}$. For the highest pressures no significant difference was seen between those velocities: turbulent noise is sufficient to induce intermittent motion below $u_{t}$ and to initiate transport, and the saturation time is rather short. For pressures below around $200$~hPa, however, the relaxation time becomes larger than the observation timescale (around a minute), leading to slightly different estimates of $u_{t}$ (less than $1$\%) when the fan rotation frequency is increased or decreased. Moreover, the turbulent noise in the transport region is not sufficient any longer to sustain intermittent transport of individual grains below $u_{t}$ observable over a minute timescale.

\textit{Error bars~--~}To determine error bars, we have performed a blind test to measure several times the threshold i.e. the cross-over from intermittent individual grains to burst regime at the same pressure. Repetability of the threshold value for $\Omega$ is within $2.5$\%. Another $2.5$\% uncertainty results from the transformation from $\Omega$ to $u_*$. This overall leads to an uncertainty on $u_t$ around $3.5$\% and twice that for the shear stress which scales as $u_t^2$. Note that the threshold is measured on a flat bed, without ripples. Moreover, negative feedback of sediment transport is negligible at the threshold (lower than $2\%$ on the stress), as the sediment flux itself is reduced to very few intermittent grains.

%_______________________
\subsection*{Measuring bed erosion rate}
\begin{figure}[p]
\begin{center}
\includegraphics[width=0.75\linewidth]{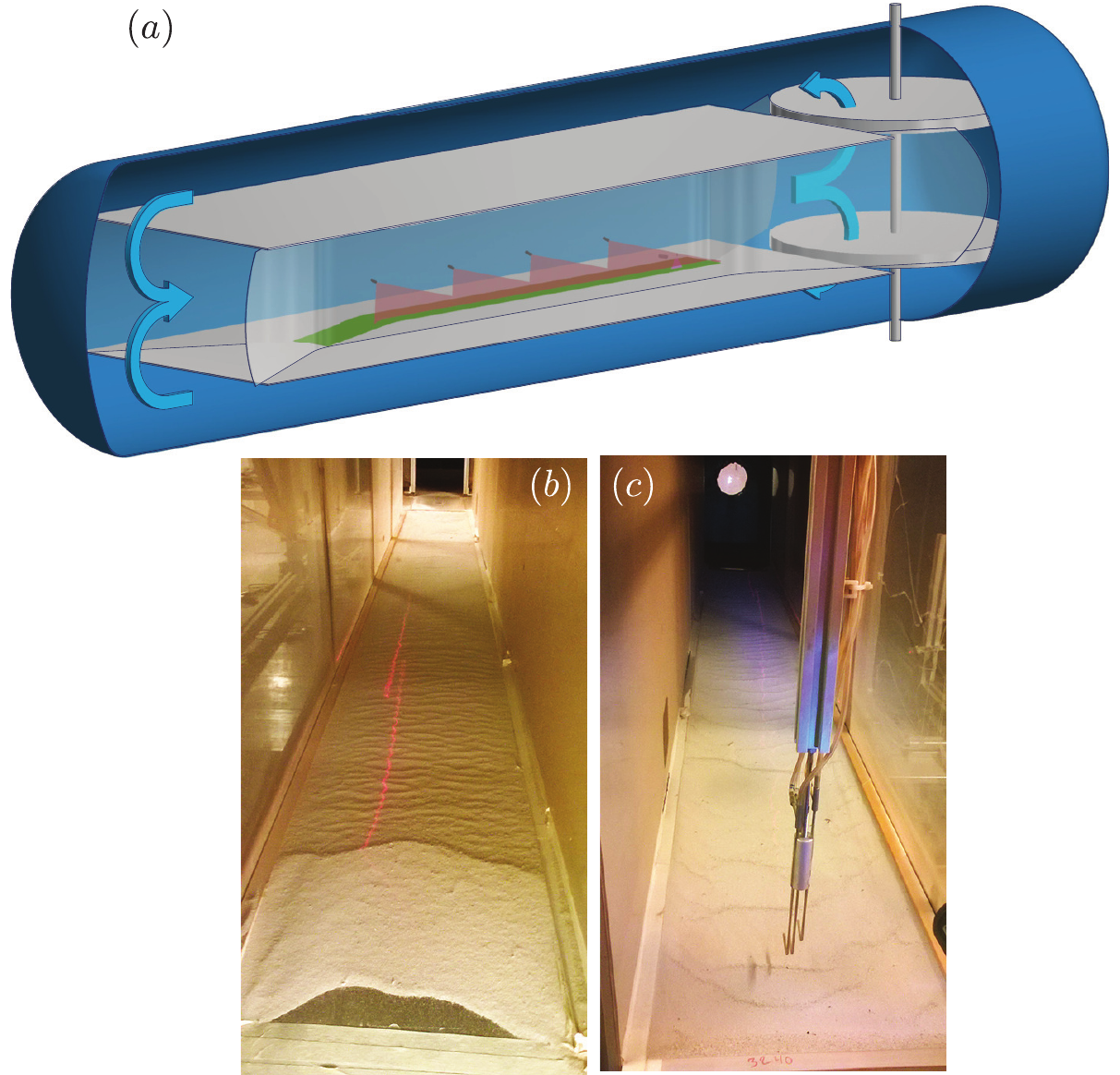}
\end{center}
\caption{(a) Schematic of insert tunnel with instruments: i) rake of pitot-static tubes connected to the HCLA-differential pressure instrument;  ii) microscope for observing near bed particle movement;  iii) lasers inclined at a low angle to the bed illuminating a line approximately along its centre; iv) web cameras recording bed topography. (b) Photograph of the laser line on the bed, looking downwind in the tunnel. (c) Photograph of the three pitot static tubes at the downwind end of the working section.} 
\label{fig:exp-setup-insert}
\end{figure}

\begin{figure}[p]
\begin{center}
\includegraphics[width=0.6\linewidth]{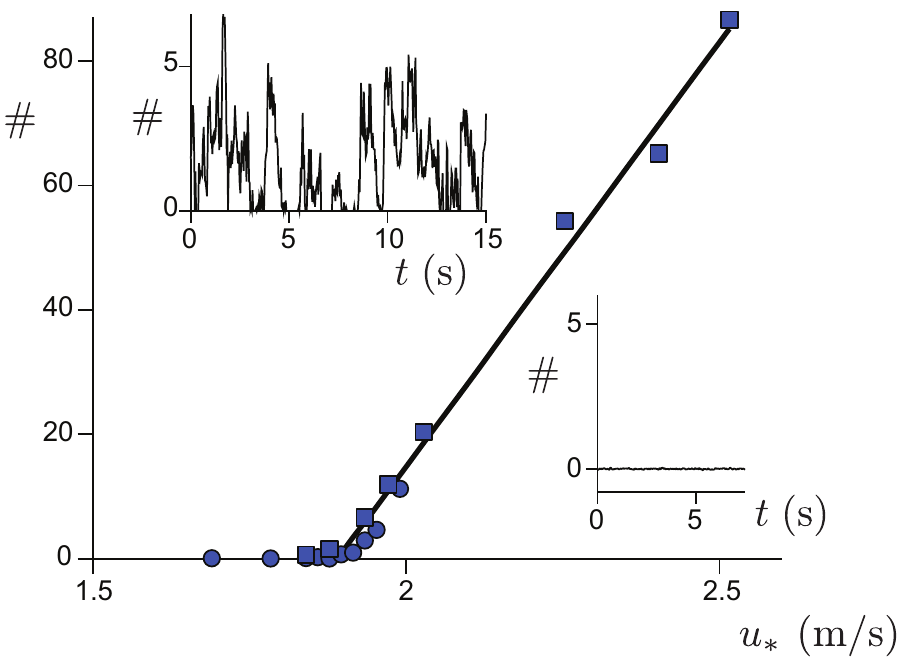}
\end{center}
\caption{Equivalent of Fig. 3a, but at low pressure $P=10^3$~Pa. Circles: data upon increasing the wind speed. Squares: data upon decreasing the wind speed. The solid line is the best linear fit trough the data. Data dispersion provides an idea of the error made due to the finite observation time and the long relaxation time. Insets: typical signals above (left) and below (right) threshold.}
\label{fig:exp-burst-lowpressure}
\end{figure}

\begin{figure}[p]
\begin{center}
\includegraphics[width=\linewidth]{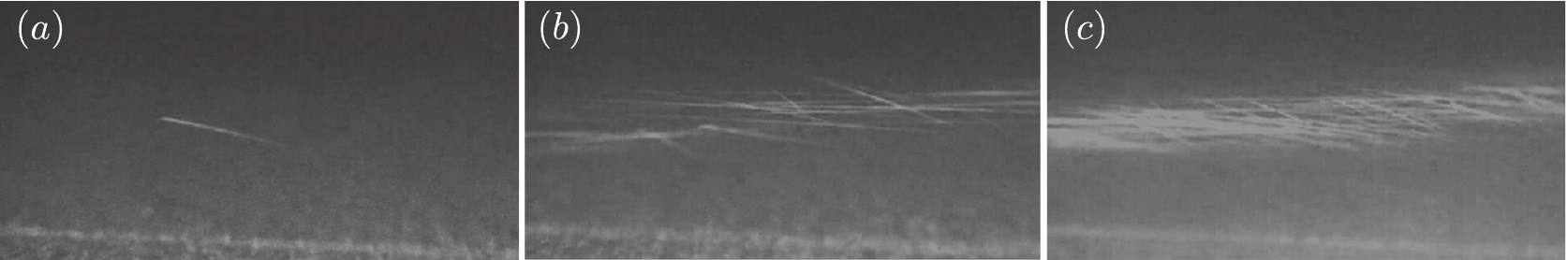}
\end{center}
\caption{Visualisation of sediment transport at ambient pressure. Microscope videogram in the regime where (a) individual grains are intermittently transported, (b) in the burst regime and (c) in the permanent transport regime. This allows us to measure the sediment transport threshold, defined as the transition from (a) to (b) with an unprecedented precision. The wind flows from right to left.}
\label{fig:exp-videogram}
\end{figure}

Four overhead web-cameras were mounted equidistantly between the inlet and outlet of the working section while four sheet lasers were placed outside the tunnel impinging the granular bed close to the center of the tunnel and forming a straight line on an initially flat bed (Supp. Fig.~\ref{fig:exp-setup-insert}). The inclination of each laser sheet was adjusted to be around $15^\circ$ from the horizontal. The positions of the laser lines were adjusted so that there was no overlap between any two lines. Changes in bed elevation resulted in a sideward displacement of the laser line, which then was recorded by the web-cameras. A four-step vertical calibration target was placed on the bed below each camera and imaged in order to verify the conversion between transverse displacement and bed elevation.

\textit{Error bars~--~}This system allowed continuous monitoring of the bed elevation profile with a resolution on the order of $0.08$~mm (one pixel). The average erosion rate has been measured at different locations, showing a statistical dispersion around $0.005$~m/s that we have used as error bars in Fig.~1b.

%_______________________
\subsection*{Measuring grain bulk density}
Having precise values for the grain bulk mass density $\rho_p$ is of prime importance to accurately compute the relevant density ratio and the other dimensionless numbers. Here, it was measured with a pycnometer of precisely known volume ($V_0 = 10.164$~cm$^3$). We first fill it with a liquid (both water and ethanol have been used) and measure its mass $m_f^0$ (in all measurements, the mass of the empty pycnometer is tared out). In a second step, we put some dry grains in the empty bottle, and measure the corresponding mass of grains $m_p$. We then fill the pycnometer with the liquid and measure the corresponding total mass of the grain / fluid mixture $m_t$. Care has been taken to avoid any air bubble in the mixture, and a vacuum pump has been used.

From volume and mass conservation equations, we obtain:
\begin{equation}
\frac{\rho_p}{\rho_f} = \frac{m_p}{m_p+m_f^0-m_t}.
\label{DensityRatioPycnometer}
\end{equation}
Knowing the fluid mass density by $\rho_f = m_f^0/V_0$, we can then deduce $\rho_p$. Values for the quartz grains we have used in the wind tunnel, as well as for the walnut shells used in the NASA Martian wind tunnel are displayed in Tab.~\ref{Tab:BulkDensityGrains}.

\textit{Error bars~--~}$\rho_p$ is measured within $1$ \%, which includes both statistical and absolute error bars.

%______________________________
\section*{Pressure and air flow calibration}

%_______________________
\subsection*{Absolute pressure}
For better accuracy, two types of gauges were employed for determining the chamber pressure: a Pirani type gauge (Pfeiffer TPR280) and a capacitance type gauge (Pfeiffer APR250). The Pirani sensor is typically accurate for low pressures (below $10$~hPa) though becomes unreliable for gas types other than air. The capacitance sensor is insensitive to gas composition and is expected to be highly accurate above $100$~hPa, though inaccurate at the lowest pressures.

\begin{figure}[p]
\begin{center}
\includegraphics[width=0.9\linewidth]{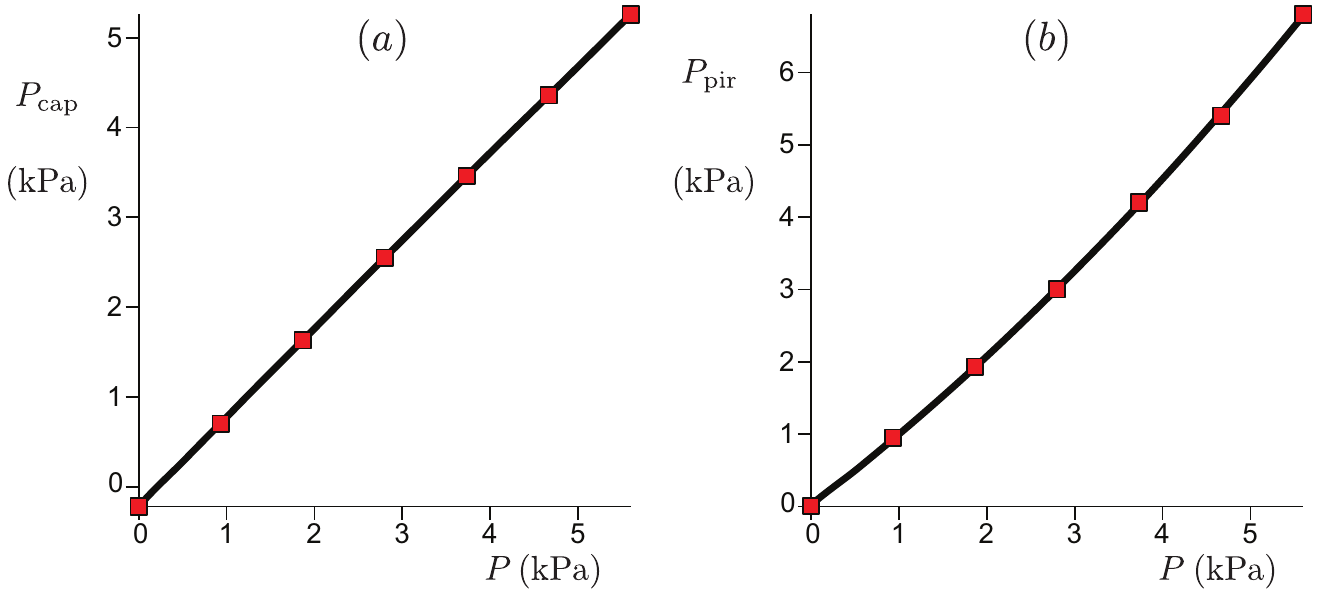}
\end{center}
\caption{Calibration data for capacitance (a) and Pirani (b) pressure sensors. Raw signal of the sensors as a function of the true pressure $P$. The capacitance sensor presents an offset value while the Pirani sensor is more accurate at small pressure.}
\label{fig:exp-PressureCalibration}
\end{figure}

The absolute calibration of these pressure sensors was performed as follows. A glass flask of precisely known volume $196.2 \pm 0.1$~cm$^3$ was filled with $1000$~hPa of air (which could be determined to an accuracy $<0.2$\% using a capacitance sensor and also calibrated with respect to absolute room pressure). This flask was then used in order to inject a known mass of air ($0.23$~g) into a larger vacuum chamber of volume $20750 \pm 10$~cm$^3$). By repeated injections of this mass of air the pressure within the vacuum chamber could be sequentially increased, in increments of $9.33$~hPa. Simultaneously measuring the pressure of this vacuum chamber using both capacitance and Pirani sensors allowed their calibration with respect to absolute pressure up to around $56$~hPa. These data are plotted in Supp. Fig.~\ref{fig:exp-PressureCalibration}. 

As can be seen, the Pirani sensor is in good agreement with absolute pressure below $10$~hPa (to better than $2$\%). The capacitance sensor has a constant offset of $2.2$~hPa with an uncertainty of $< 2$\%. Extrapolating this offset for the capacitance sensor is in good agreement with the offset observed at around $1000$~hPa supporting the assumption of good linearity for this sensor over the entire range ($1$--$1000$~hPa) \cite{JMI2019}. The measured pressure values agreed well with the manufacturer's claims of $2$\% accuracy and $0.5$\% linearity for the capacitance sensor.

The room temperature was monitored using a pt100 thermistor and was seen to be around $21 \pm 1^\circ$C. Prior to wind flow this was taken as the air temperature within the chamber. During prolonged high wind flow, a small increase in chamber pressure (of a few \%) was observed, which is consistent with the expected heating of the air by friction and does not affect the gas density.

\textit{Error bars~--~}Combining the measurements of both calibrated sensors, general uncertainties in absolute pressure and density $\rho_f$ are less than $2$ \% although they level off at $10~\rm{Pa}$ below $500~\rm{Pa}$. Horizontal error bars are therefore much smaller than symbol size in figures 1a, 1b, 3b and 4.

%_______________________
\subsection*{Calibration of Pitot tubes}
Measurement of differential pressure from up to five pitot-static tubes was made using miniature amplified pressure transducers (First Sensor A/G HCLA 02X5) which are insensitive to gas composition. For this transducer it's sensitivity (G) is linear over the 2.5 hPa full range while it's offset (O) for zero pressure varies between transducers. At low ambient pressure $P$, the differential pressure of a pitot tube is very small and we have no information on the stability of the electronics controlling recording and A/D conversion. A calibration experiment was made in order to estimate G and O at varying pressures ($3.4$, $14$, $37$, $110$, $308$ and $972$~hPa). In the experiment, the dynamic and static ports of the pressure transducers were connected, respectively, to one of two pressure chambers connected via a valve. Each chamber was supplied with a Pirani and a capacitance sensor and one chamber was connected to a vacuum pump. When open, the valve between the chambers secured the same pressure in both of them. When closed, a valve (to the ambient) in the second chamber allowed small amounts of air to enter, creating a pressure difference between the chambers, in small steps up to the $2.5$~hPa full range. The calibration verified that output was truly linear and G independent of pressure while O for each transducer was sensitive to pressure as well as temperature. The average G for each of the five calibrated transducers is given in Supp. Tab.~\ref{Tab:CalibDiffPressureData}. The calibration showed that before each experiment it is important to record the offset value for each sensor during 2-3 minutes with stable temperature in order to estimate O at the actual pressure.

\begin{figure}[p]
\begin{center}
\includegraphics{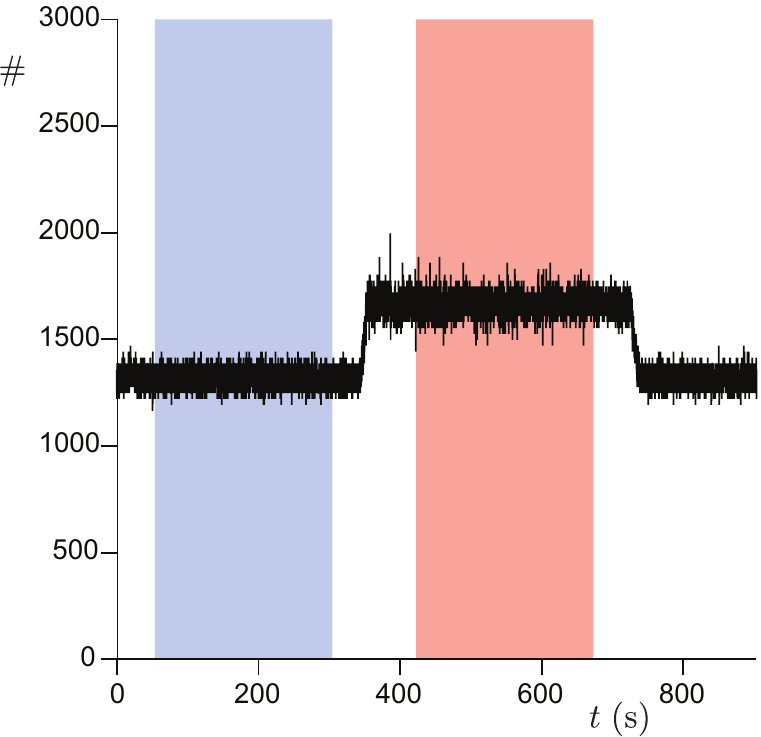}
\end{center}
\caption{Raw data (counts) for a $2.5$~hPa pressure sensor measuring differential pressures from a pitot-static tubes at $83$~mm height above the granular bed below transport threshold. The blue and red hatched areas mark the beginning and end of the intervals during which the average flow speed was measured.} 
\label{fig:ExPressureRecord}
\end{figure}

\begin{figure}[p]
\begin{center}
\includegraphics{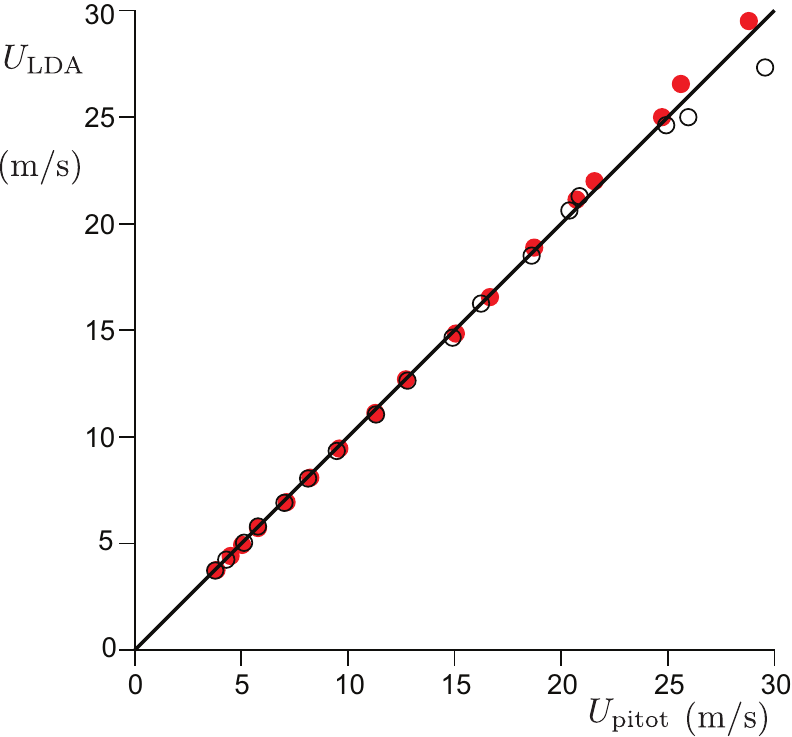}
\end{center}
\caption{Air velocity above the granular bed measured with a LDA-instrument plotted versus velocity measured with a pitot static tube connected to the HCLA-instrument. The two symbols represent two data sets.}
\label{fig:LDAvsPitot}
\end{figure}

When the average pressure difference is based on readings at $50$~Hz over a 3-minute interval, the error of the difference is on the order of $4$\% and the corresponding error on a velocity of about $20$~m/s at Martian pressure is less than $2$\%. An example for the output recorded at $11$~hPa is shown in Supp. Fig.~\ref{fig:ExPressureRecord}. In an attempt to investigate further the quality of velocity data based on pitot tube data, we have placed one tube at the end of the working section at a $83$~mm elevation above the bed and recorded the pressure difference as described above. Outside one of the windows in the environmental wind tunnel a 2-D Dantec Flow-Lite laser-Doppler anemometer (LDA) instrument was set up with beams crossing $10$~mm upwind of the tip of the pitot tube. Seeding dust particles was made through a valve placed at the center of the upwind end of the environmental wind tunnel. Measurements were performed at combinations of $P$ (above $3.4$~hPa) and $\Omega$ for which the saltation threshold was estimated. Two sets of data are available, displayed in Supp. Fig.~\ref{fig:LDAvsPitot}.

LDA measurements at the lowest pressures ($3.4$ and $6.6$~hPa) were based on a few dust particles only, corresponding to very short intervals immediately after injection. The deduced speeds may then not be representative of the average fluid velocity over the entire recording period. At all higher pressures, we recorded particle speed during several and longer intervals distributed over the entire recording period. The minimum number of counts was 350, and in most cases, the number was much higher and often higher than $1000$ particles. As can be seen from the Supp. Fig.~\ref{fig:LDAvsPitot}, air velocity can be measured satisfactorily using the pitot-static tube/HCLA set-up from ambient to below Martian pressures.

%_______________________
\subsection*{Measuring wind velocity profiles}
During the threshold experiment three pitot-static tubes (outer diameter 4 mm) were placed at the downwind end of the working section with their tips at elevations of $10$, $20$ and $40$~mm above the granular bed. At ambient pressure, when a pitot-static tube is placed in the saltation layer, a stagnation bubble forms in front of the orifice of the dynamic tube, which prevents grains from entering the tube. However, at decreasing fluid pressure, the stagnation bubble weakens and saltating grains eventually accumulate in the tube and gradually clog it, even if those moving grains are not numerous when working close to the transport threshold. Therefore, for those low pressures, only a few reliable wind velocities were recorded at one or more elevations following the start of an experimental run.

We have consequently made a series of post-experiment velocity calibrations and tests. Here a `fixed-bed' was prepared by gluing a sample of the grains used in the experiment to a metal bed. We also added pitot tubes at $5$ and $15$~mm elevations thus enabling the recording of the velocity profile based on five elevations above the bed. Once the pitot tube rake was installed in the wind tunnel, a precision elevation gauge ($\pm 0.1$~mm) was used to measure the exact elevation above the bed of each pitot tube. For each values of $P$ and $\Omega$ for which the saltation threshold for the granular bed was estimated, we sampled a velocity at each elevation of a pitot tube. After recording was made for pressures the pitot-rake was lifted by $2.5$~mm and a second set of velocities recorded at the new elevations. At every pressure we can thus construct a velocity profile based on measurements at ten different elevations (Supp. Fig.~\ref{fig:WindProfilesForShearVelocity}). These data must be fitted by a hydrodynamic model to deduce the corresponding wind shear velocity $u_*$, as described below.

Experimental runs for the measurement of the threshold were performed with a bed initially made uniformly flat and as close to threshold only marginal transport occurs, no ripples developed.  It is then justified to assume that, being at the onset of saltation, the shear stress measured above a bed of loose particles is similar to that above a fixed surface composed of the same (glued) grains. Moreover, we have gone through the differential pressure data measured above the granular bed and picked the runs that we think contains data at two or three heights that are not (or only to a small degree) influenced from clogging of the dynamic tube. For the same pressure, we have calculated the ratio between the speed measured at an almost similar height (deviation less than $1$~mm) above the granular bed and the fixed bed (Supp. Fig.~\ref{fig:UFixedGranularBed}). For most pressures the ratio is close to unity, while at the lowest pressure of $11$~hPa one of the pitot tubes is obviously partly clogged. We are then confident that the threshold shear stress values derived from the velocity profiles measured on the fixed bed are representative of those on the granular bed. We emphasize that sediment transport at the threshold --~as defined here~-- is limited to intermittently moving grains, whose negative feedback on the wind is negligible ($2\%$ on the shear stress, at most).

%_______________________
\subsection*{Measuring the wind shear velocity}
In order to convert the frequency $\Omega$ into a wind shear velocity $u_*$, we need to fit the velocity profile by a turbulent model, which we adapt here from the literature. We consider a fluid flow along the $x$ direction over a flat bed. Here, $z$ is the crosswise axis normal to the bed, and $y$ is spanwise. Following the standard separation between average quantities and fluctuating ones (denoted by a prime), the equations governing the mean velocity field $u_i$ and the pressure $p$ can be written as
\begin{equation}
\partial_i u_i  = 0
\qquad {\rm and} \qquad
\rho_f \partial_t u_i+u_j \partial_j u_i  = \partial_j \tau_{ij}-\partial_i p,
\label{NS}
\end{equation}
where $\tau_{ij}$ contains the Reynolds stress tensor $-\rho \overline{u'_i u'_j}$. We use a first-order turbulence closure to relate the stress to the velocity gradient. It involves a turbulent viscosity resulting from the product of a mixing length and a mixing frequency, representing the typical eddy length and time scales. The mixing length $\ell$ depends explicitly on the distance to the bed. The mixing frequency is given by the strain rate modulus, which, in the homogeneous situation along the $x$-axis assumed here, reduces to $\partial_z u_x$. Adding up viscous and turbulent contributions, the shear stress writes
\begin{equation}
\tau_{xz} = \eta \partial_z u_x + \rho_f \ell^2 |\partial_z u_x| \partial_z u_x,
\label{ShearStress}
\end{equation}
where $\eta$ is the dynamic fluid viscosity. In order to account for both smooth and rough regimes, we adopt here a van Driest-like expression for the mixing length \cite{CDA2017}:
\begin{equation}
\ell=\kappa \left[ z + rd \right] \left[1-\exp\left(- \frac{(\tau_{xz} \rho_f)^{1/2}\left[ z + sd \right]}{\eta \mathcal{R}_t}\right)\right].
\label{ellcombo}
\end{equation}
In this expression, $\kappa=0.4$ is the von K\'arm\'an constant, $d$ is the sand equivalent bed roughness size, here set to the diameter of the grains used in the experiments, and $\mathcal{R}_t$ is the van Driest transitional Reynolds number, set to $\mathcal{R}^0_t \simeq 25$ in the homogeneous case of a flat bed \cite{P2000}. The exponential term suppresses turbulent mixing within the viscous sub-layer, close enough to the bed \cite{WW1989}. $rd$ corresponds to the standard Prandtl hydrodynamical roughness extracted by extrapolating the logarithmic law of the wall at vanishing velocity. $sd$ controls the reduction of the viscous layer thickness upon increasing the bed roughness. The dimensionless numbers $r=1/30$ and $s=1/3$ are calibrated with measurements of velocity profiles over varied rough walls \cite{SF2009,FS2010}. We have checked that uncertainties on these values only affect the results in a negligible way, as our experiments take place in the smooth aerodynamical regime.

In order to account for the tunnel geometry, we assume a shear stress profile of the form:
\begin{equation}
\tau_{xz} = \rho u_*^2 \exp\left[- \left(\frac{z}{\delta}\right)^2\right],
\label{ShearStressInTunnel}
\end{equation}
where $u_*$ is the shear velocity and $\delta$ is the thickness of the turbulent boundary layer. $\rho_f u_*^2$ is the shear stress that the air flow applies to the granular bed. We compute corresponding the velocity profile $u_x(z)$ with a numerical integration of Eqs. (\ref{ShearStress}-\ref{ShearStressInTunnel}) coupled together.

\begin{figure}[p]
\begin{center}
\includegraphics[width=0.9\linewidth]{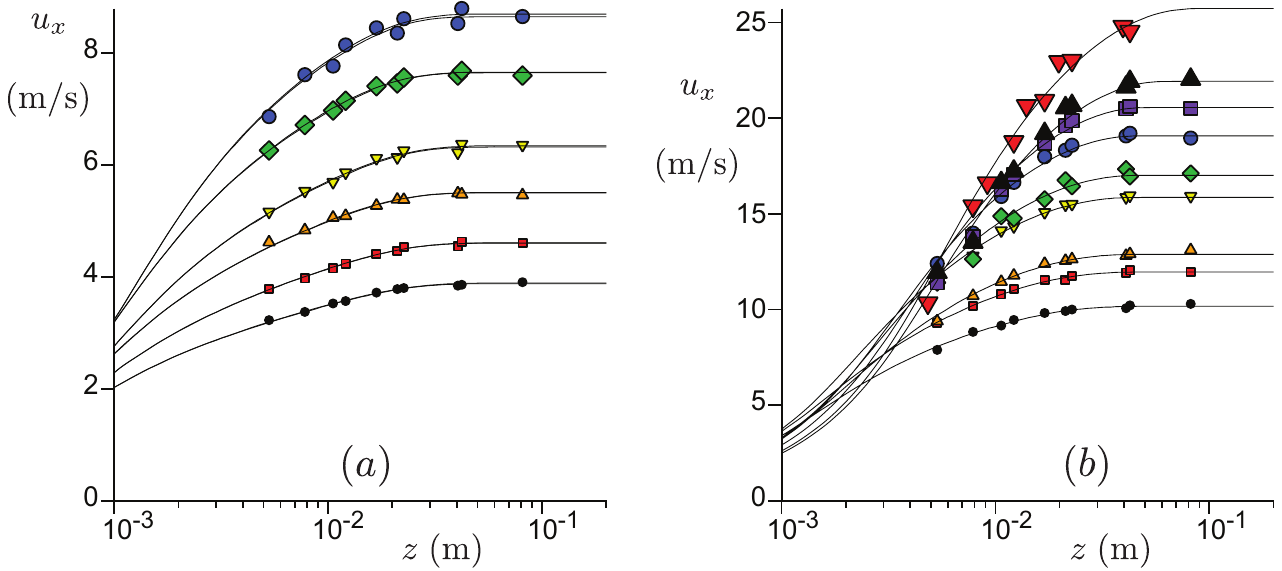}
\end{center}
\caption{Wind velocity profiles at different pressures. The error bars are smaller than the symbol size. Solid lines: theoretical fits (see section 2.4), from which the corresponding wind shear velocity $u_*$ is deduced. Symbols in panel (a) from bottom to top: $P=998$, $P=720.2$, $P=518.3$, $P=373.9$, $P=269.8$ and $P=194.6$ hPa. Symbols in panel (b) from bottom to top: $P=140.9$, $P=102.1$, $P=74.3$, $P=54.1$, $P=39.05$, $P=29.1$, $P=21.7$, $P=16.1$ and $P=13.4$ hPa. Uncertainties are on the order of the symbol size or smaller.}
\label{fig:WindProfilesForShearVelocity}
\end{figure}

\begin{figure}[p]
\begin{center}
\includegraphics{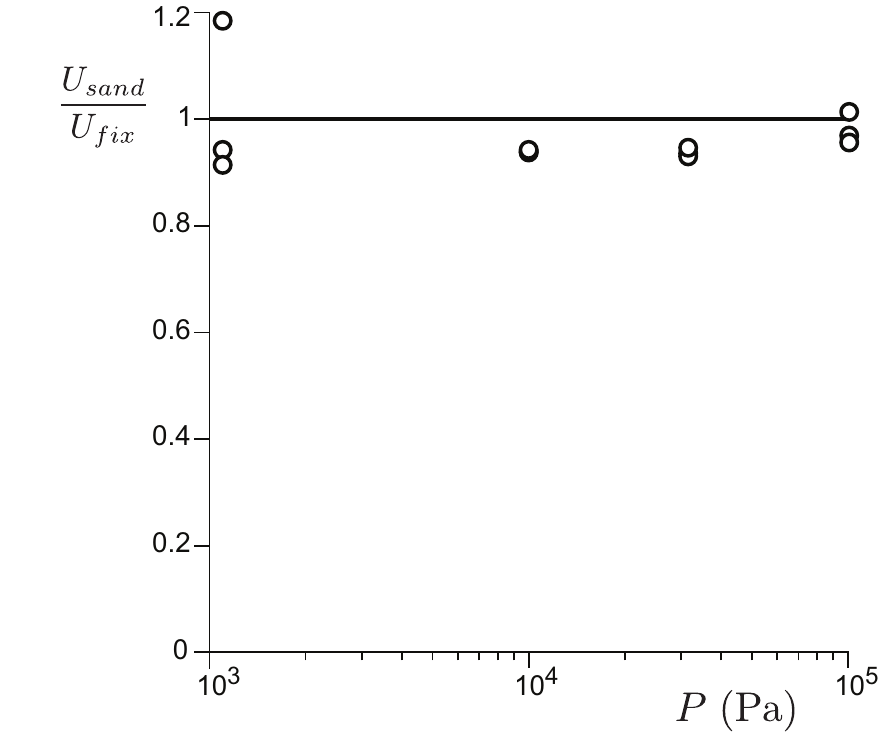}
\end{center}
\caption{Comparison of velocities measured over a fix ($U_{\rm fix}$) or a mobile ($U_{\rm mob}$) granular bed, for different pressures, at the threshold value $\Omega=\Omega_t$. Although sediment transport at the threshold is residual and intermittent, it may gradually clog the Pitot tubes leading to a slightly smaller apparent velocity. The clogging problem is enhanced at pressures lower than $10^3~\rm{Pa}$. Altogether, the deviations to unity are consistent with the dispersion of data points, providing direct proof that negative feedback of sediment transport on the wind velocity is negligible at threshold.}
\label{fig:UFixedGranularBed}
\end{figure}

\begin{figure}[p]
\begin{center}
\includegraphics[width=0.9\linewidth]{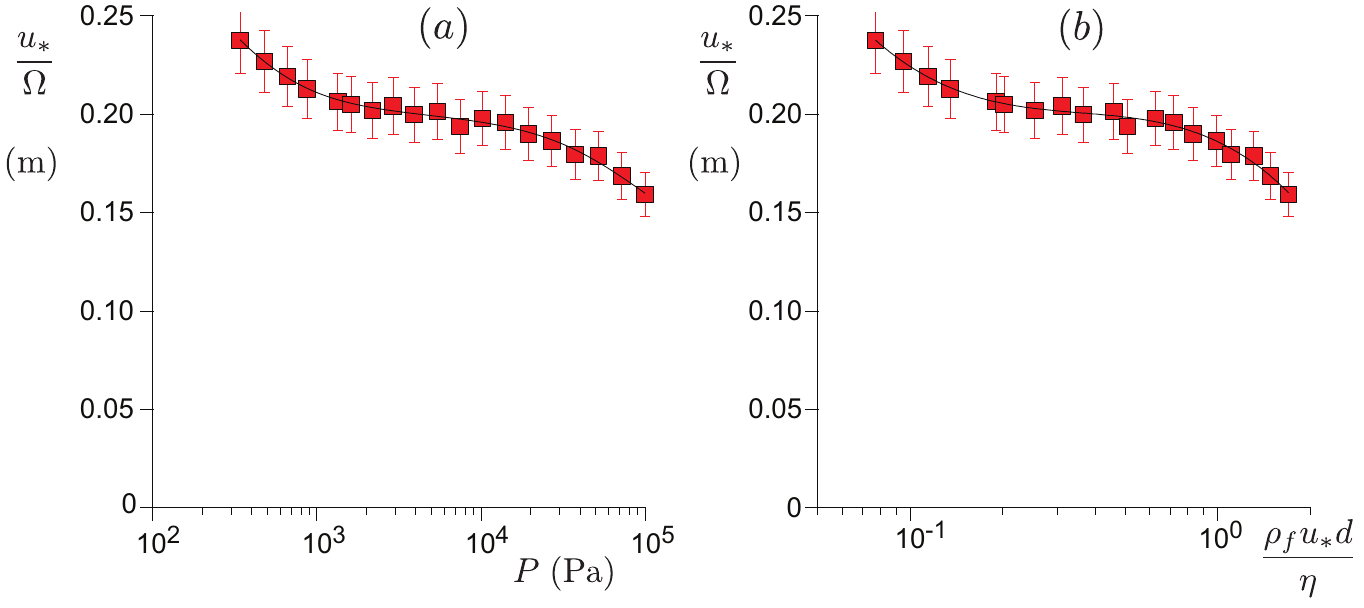}
\end{center}
\caption{The ratio of friction speed $u_*$ and fan frequency $\Omega$ (in Hz) as a function of the tunnel pressure (a) and Reynolds number (b). The solid line represents an empirical fifth order polynomial fit.}
\label{fig:expustarsrpm}
\end{figure}

Adjusting both $u_*$ and $\delta$, we can reproduce the velocity data described in the previous section, see Supp. Fig.~\ref{fig:WindProfilesForShearVelocity}. The wind profiles were measured close to the transport threshold. In Supp. Fig.~\ref{fig:expustarsrpm}a we display the corresponding friction speed $u_*$ rescaled by $\Omega$ as a function of the pressure $P$. The ratio $u_*/\Omega$ varies slowly with $P$. The thickness $\delta$ also gently varies with $P$, increasing from $2$ to $6$~cm when decreasing the pressure from the ambient to a few hPa. These values corresponds to the altitude $z$ at which the velocity profiles quit the boundary layer and saturate (Supp. Fig.~\ref{fig:WindProfilesForShearVelocity}).

The true control parameter is in fact not the pressure, but the Reynolds number based on the grain diameter $d$, which depends on $u_*$ and on the fluid density $\rho_f$. In practice, in order to interpolate between points, and to include data points associated with experimental runs in CO$_2$, we have used an empirical fifth order polynomial fit of this calibration curve $u_*/\Omega$ as a function of $\rho_f u_* d/\eta$ (Supp. Fig.~\ref{fig:expustarsrpm}b).

The best fit of the velocity profiles by the model equations leads to residuals that are consistent with the velocity measurement accuracy. The model depends on two fitted parameters: the shear velocity $u_*$ and the thickness of the turbulent boundary layer $\delta$. We emphasize  that the experiments are performed with $125\rm{\mu m}$ grains so that the grain based Reynolds number is always smaller than $2$ --~value at ambiant pressure. The calibration runs on a fixed bed are therefore all in the smooth aerodynamic regime, where the viscous sublayer is significantly larger than the grain size. Including both the statistical uncertainties, defined from the residuals and the instrumental resolution, the uncertainty on the measurement of $u_*$ from a velocity profile is around $7\%$ over the whole pressure range. The uncertainty on $\delta$ decreases logarithmically with the pressure, from $7\%$ at $P=10^2$~Pa to $0.4\%$ at at $P=10^5$~Pa. Using the calibrated relation between $\Omega$ and $u_*$, a factor $3$ can be gained on the accuracy on $u_*$.

\textit{Error bars~--~}The final error bars for the value of the determination of $u_*$ is obtained from the law relating the ratio $u_*/\Omega$ to the Reynolds number $\rho_f u_* d/\eta$. The systematic error can be decomposed into two parts: the miscalibration as a whole (all values, regardless the pressure, would be over or under estimated) is smaller than $1$\%. The systematic error associated with the slope of the relation may also produce systematic variations with pressure, at worst on the order of $2.5$\%.

%______________________________
\section*{Sediment transport thresholds from the experimental and theoretical perspectives}
We discuss in this section possible ambiguities in the naming of sediment transport thresholds. On the one hand, they are defined in wind tunnel and field experiments, based on pragmatic measurement procedures. On the other hand, they are defined in the theoretical and numerical literature based on conceptual ideas and simplified descriptions. While both are equally legitimate, we argue that these two definitions do not perfectly coincide. Our purpose is of course not to introduce an epistemic hierarchy between theory, wind tunnel experiments and field, but rather to warn the reader on the sense in which concepts are used here.

From pioneering works of Bagnold \cite{B1941}, two distinct thresholds for sediment transport have been standardly introduced in the literature: the static or fluid threshold and the dynamic or impact threshold. These two thresholds have clear definitions in theoretical models. The static threshold characterizes the minimal flow shear velocity needed to entrain the surface grains from a static bed. It results from a balance of the forces acting on such a grain: weight, friction with the bed, drag, and possibly cohesion (see e.g. \cite{SY2000, CA2006}). By contrast, it is argued that sediment transport can be sustained below that value, due to impacts and rebounds of the grains with the bed, down to the dynamic threshold, which must then be computed taking into account the grain trajectories (see e.g. \cite{A2004, CA2006, K2010, DCA2011, PD2018, PCVD2020}). Importantly, in their theoretical acceptations, the static threshold shear stress is larger than the dynamic threshold shear stress. Note that most models ignore the effect of intrinsic wind fluctuations and bed heterogeneities.

Over several decades, sediment transport thresholds were measured experimentally in aeolian \cite{C1945, H1971, GWLIP1976, GLWIP1980, IW1982, RIR1996, RS1999, CRH2015, MK2018, SSE2020}, subaqueous \cite{FLvB1976, YK1979, BM1997, LR1998, NLG2003, GMM2010} or intermediate \cite{BBMSWE2015,BSRNKSB2020} conditions. In those experiments, the dynamic or cessation threshold is defined in most papers as the extrapolation to vanishing flux of the relation between the average sediment flux and the imposed constant, shear velocity $u_*$. This threshold is accurately defined, as it results from the fitting of robust quantities (sediment flux or related proxies). By contrast, more variation is found in the definition of the static threshold in experimental papers. The idea is to determine the moment when first isolated grains are set in motion upon increasing the wind speed: these grains can be detected as just detached (rocking/rolling), or entrained (leaving the bed and bouncing), or associated with sporadic saltation. As a consequence of this arbitrariness in the criterion for the incipient grain motion, those measurements generally present huge data dispersion. In the field, the shear stress presents besides fluctuations at the relevant time-scale for sediment transport, which makes the link between wind tunnel experiments and field measurement complex --~and a matter of ongoing debate. In this context, it has been proposed that aeolian saltation in the field, for a given mean wind velocity, is both sensitive to fluid and impact thresholds (see e.g. \cite{MK2018}).

Here, we propose the following criterion to identify the static threshold (in the theoretical sense) in experiments. As, by definition, incipient motion does not depend on the grain inertia, the curve relating the Shields number $\Theta$ to the Galileo number $\mathcal{G}$ is independent of the density ratio $\rho_p/\rho_f$, and is thus identical in aeolian and in subaqueous conditions. All measurements should therefore collapse on that curve. However, published data in the literature reporting static thresholds (in the experimental sense) for sediment transport in air (see e.g. \cite{C1945, GLWIP1980, RIR1996, RS1999, SSE2020}) are typically below the static threshold measured under water (Fig.~4). The conclusion is therefore that those values of saltation thresholds reported in the experimental and field literature are not static thresholds, as defined by theory. Note that the wind shear velocity at which the first granular motion is observed would be even lower than the transport threshold reported here and would therefore be even less comparable to the theoretical static threshold.

In this work, we have considerably reduced error-bars on the transport threshold by using a specific experimental procedure. We determined quantitatively the cross-over between a regime of residual sediment transport, composed by intermittent, individual moving grains, and a permanent though bursty transport regime. Uncertainties are small as this quantity is robust with respect to turbulent fluctuations. Increasing the wind speed from a static bed, or decreasing it from sustained transport, the same threshold shear velocity is measured, within a few percent. This may be interpreted as an effect of turbulent fluctuations, which play the role of thermal activation in bifurcation theory: we do not observe any significant hysteresis, reason for which we name it `the' threshold, here. As a consequence, we do not name it `static' nor `dynamic' but reserve these words to their theoretical meanings. Note that we use here a closed wind tunnel experiment, so that transported grains may be reinjected from time to time at the entrance of the working section.

%______________________________
\section*{Static threshold model}
In this section, we summarise the modelling of the threshold shear velocity for the incipient motion of a grain at the surface of the bed (static threshold in the theoretical sense). We do not develop here a new theoretical framework but gather elements from previously published studies, in particular from Refs.~\cite{CA2006,DCA2011,CDA2017}. We also provide values for the various empirical factors or parameters, obtained by calibration and best fit of experimental data independent of the present study.

%_______________________
\subsection*{Drag force}
We take the drag force ${\vec f}_d$ of the flow on a grain as a function of the difference between the grain velocity $\vec v$ and the fluid velocity $\vec u$ at the grain's location:
\begin{equation}
{\vec f}_d = \frac{\pi}{8} \rho_f d^2 C_d |{\vec u} - {\vec v}| ({\vec u} - {\vec v})
\label{dragforce}
\end{equation}
where the drag coefficient $C_d$ depends on the particle's Reynolds number based on that velocity difference $\mathcal{R}_u = \rho_f |{\vec u} - {\vec v}| d/\eta$ and is empirically written in the form of two terms to account both inertial and viscous (Stokes) contributions:
\begin{equation}
C_d = \left(C_\infty^{1/2}+A / \mathcal{R}_u^{1/2} \right)^2.
\label{dragcoeff}
\end{equation}
The best fit of the drag curve measured for natural sand grains gives $C_\infty \simeq 1$ and $A \simeq 5$ \cite{FC2004}.

%_______________________
\subsection*{Static force balance}
We hypothesise that the static threshold $u_{ts}$ results from the force balance on a grain at rest at the bed surface. Along the flow direction, the drag force from the fluid flow is opposed to the effective bed friction. Following the above expression of the drag force, this balance writes:
\begin{equation}
\alpha \frac{\pi}{8} \rho_f d^2 C_d u^2 = \frac{\pi}{8} \mu (\rho_p - \rho_f)gd^3,
\label{forcebalance}
\end{equation}
where we have introduced the friction coefficient $\mu=0.6$ and a factor $\alpha=1/2$ to account for the fact that the only the upper half of the grain is submitted to the hydrodynamic stress. $u$ is the flow velocity at the grain's location $u=u_x(\beta d)$.

To get the corresponding shear velocity $u_*=u_{ts}$, the computation of the whole profile $u_x(z)$ is needed. This is obtained by integration of the hydrodynamic horizontal momentum balance, which reduces to $\tau_{xz} = \rho_f u_*^2$, associated with its expression with the velocity gradient (\ref{ShearStress}) and the mixing length (\ref{ellcombo}). Defining $u_x(z) \equiv u_* \mathcal{U}(z/d)$, one then needs to integrate the differential equation
\begin{equation}
\Upsilon^2 |\mathcal{U}'|\mathcal{U}'+ {\mathcal R}^{-1} \mathcal{U}'=1,
\qquad {\rm or\ equivalently} \qquad
\mathcal{U}'=\frac{-1 +\sqrt{1+4\Upsilon^2 {\mathcal R}^2}}{2\Upsilon^2 {\mathcal R}}\,.
\label{EqDiffmathcalU}
\end{equation}
with the boundary condition $\mathcal{U}(0)=0$ corresponding to the no-slip condition of the wind at the solid interface. In the above equation the mixing length is made dimensionless by
\begin{equation}
\ell/d \equiv \Upsilon(\zeta)= \kappa (\zeta+r)\,(1-\exp(-{\mathcal R}(\zeta+s)/\mathcal{R}_t^0)).
\label{defUpsilon}
\end{equation}
With these notations, the force balance can be rewritten as:
\begin{equation}
C_\infty^{1/2}  \mathcal{R}{\mathcal{U}(\beta)} + A \left[ \mathcal{R}{\mathcal{U}(\beta)} \right]^{1/2} = \left(\frac{4\mu}{3\alpha}\right)^{1/2} \,\mathcal{G},
\label{forcebalanceadim}
\end{equation}
where we have introduced the Reynolds number
\begin{equation}
\mathcal{R} = \frac{\rho_f u_* d}{\eta}
\label{Reynolds}
\end{equation}
and the Galileo number
\begin{equation}
\mathcal{G} = \frac{1}{\eta} \sqrt{\rho_f (\rho_p-\rho_f)gd^3}.
\label{Galileo}
\end{equation}
This allows us to deduce $\mathcal{U}(\beta)$ as a function of ${\mathcal R}$, and to compute the static condition.  The best fit to the subaqueous threshold data gives $\beta=0.85$ (Fig. 4).

Rather than the shear velocity or the Reynolds number, it is traditional in the context of sediment transport to work with the Shields number
\begin{equation}
\Theta = \frac{\rho_f u_*^2}{(\rho_p-\rho_f)gd} = \left( \frac{\mathcal{R}}{\mathcal{G}} \right)^2.
\label{Shields}
\end{equation}
The static condition thus relates $\Theta$ to $\mathcal{G}$, independent of $\rho_p/\rho_f$. Any dependence on the density ratio is the signature that the threshold is of dynamical origin (see next section).

Finally, following \cite{CA2006}, to take into account cohesion between the grains, a correction is added through a modified gravity acceleration:
\begin{equation}
g^*=g \left[1+\frac{3}{2}\;\left(\frac{d_m}{d}\right)^{5/3}\right].
\label{cohesion}
\end{equation}
Here, we have obtained the best data fit of the subaqueous threshold data with $d_m=6$~$\mu$m. This length scale can be interpreted as the particle size below which Van der Waals adhesive contact forces become dominant over the particle's weight.

%______________________________
\section*{Dynamic threshold model}
We summarise in this section a model of dynamic threshold shear velocity (in the theoretical sense). As in the previous section, this summary gathers elements from previously published works \cite{A2004,CA2006,DCA2011,CDA2017}, and provides values for the various empirical parameters. We assume that the complexity of transport can be modelled by a single type of trajectories.

%_______________________
\subsection*{Trajectory integration}
Starting from the Newton's second law of motion, the dynamical equation for the grain velocity $\vec v$ is:
\begin{equation}
\frac{1}{6}\pi d^3 \rho_p \frac{d \vec v}{dt}= \frac{\pi}{8} C_d d^2 \rho_f |\vec u-\vec v| (\vec u-\vec v)+\frac{\pi}{6} (\rho_p - \rho_f) \vec g d^3
\label{DynamicEquaMotion}
\end{equation}
In a dimensionless form where velocities are rescaled by $u_*$, lengths by $d$ and time by $1/(du_*)$, one obtains:
\begin{equation}
\frac{d \vec v_a}{dt_a} = \frac{3}{4} C_d  \frac{\rho_f}{\rho_p} |\vec u_a-\vec v_a| (\vec u_a-\vec v_a)- \frac{\rho_f}{\rho_p} \left( \frac{\mathcal{G}}{\mathcal{R}} \right)^2 \vec e_z,
 \label{DynamicEquaMotionRescaled}
\end{equation}
with the rescaled drag coefficient (see Eq.~\ref{dragcoeff})
\begin{equation}
C_d = \left(C_\infty^{1/2}+A \frac{1}{\left(\mathcal{R} |\vec u_a-\vec v_a| \right)^{1/2}} \right)^2,
\label{dragcoeffRescaled}
\end{equation}
where $u_a = u/u_*$, $v_a = v/u_*$ and $t_a = t d u_*$. The wind velocity profile $u(z)$ is computed as in the previous section (Eq. \ref{EqDiffmathcalU}). In this simplified description, the typical grain trajectory is assumed to present an initial angle $\theta_0$. We also hypothesise that the rebound law is controlled by a velocity ratio $e=v_\uparrow/v_\downarrow$, which depends on the grain Stokes number according to \cite{GLP2002}:
\begin{equation}
{\rm St} = \rho_p v_\downarrow d/\eta = \frac{\rho_p}{\rho_f} \frac{v_\downarrow d}{u_*} {\mathcal{R}}.
\label{StokesNumber}
\end{equation}
%

%_______________________
\subsection*{Dynamic steady state}
We express the criterion for steady transport at the dynamic threshold shear velocity $u_t$ with the condition that the kinetic energy of the typical grain trajectory is just enough to escape from the potential traps between its neighbours. This condition can be interpreted as a unit replacement capacity during rebounds. In the absence of wind, the escape velocity simply scales as $\sqrt{gd}$. Due to the wind, the trapping must reduce and vanish at the static threshold, which we express as:
\begin{equation}
v_\uparrow = a \sqrt{gd \left( 1 - \frac{u_t^2}{u_{ts}^2} \right)}.
\label{EscapeVelocity}
\end{equation}
The depth of these traps scale with the grain size. The quadratic dependence in shear velocity comes from the wind stress, which reduces the work of the grain's weight. The take off velocity $v_\uparrow$ vanishes at the static threshold $u_t = u_{ts}$, as it should. Eq.~\ref{EscapeVelocity} is an improvement with respect to earlier versions of the model \cite{A2004,CA2006,DCA2011,CDA2017}. The best fit to aeolian (ambient pressure) data gives $a=11$, $\theta_0=50^\circ$ and $e=0.36$ (Fig. 4).

%______________________________
\section*{Sediment transport and ripple scaling laws}
In this section we gather different elements from the existing literature to form a theoretical framework in which we can interpret our experimental results regarding transport, erosion and ripple formation.

%_______________________
\subsection*{Saltation fluxes}
We formulate a three-layer model of aeolian sediment transport (Fig. 4). Specific terminology (saltation, reptation, creep) has been proposed to distinguish between the different modes of transport. Their precise definitions, however, are not consistent throughout the literature. Here we refer to saltation as the generic word for aeolian grain motion, and define the specific physical processes in relation to a corresponding transport layer.

\textit{Feedback layer} --- The sediment flux is dominated by the central region of the transport layer. It is located below Bagnold's focal point \cite{B1941}, where the grains exert a negative feedback on the air flow \cite{O1964,McE1993,DCA2011,VROD2015}, and is well described by the approach of Ungar \& Haff \cite{UH1987} summarised below. Its thickness is on the order of several tens of $d$ or a centimeter \cite{Cetal2009}. We shall refer to it below as the `feedback layer'. Following these authors, for a given wind shear stress $\rho_f u_*^2$, the windward momentum balance sets the partition between the fluid-born and grain-born stress contributions:
\begin{equation}
\rho_f u_*^2 = \tau_f + \tau_p.
\label{StressPartition}
\end{equation}
In the steady state, the fluid-born stress is reduced to its threshold value $\tau_p = \rho_f u_t^2$, due to the grains'  feedback, and we then deduce $\tau_p = \rho_f (u_*^2 - u_t^2)$. Again, the complexity of transport is assumed to be modelled using a single grain trajectory. The grain-born stress can also be expressed as a function of the vertical particle flux $\varphi$ and the horizontal velocities of ascending or descending grains, $v_\uparrow$ and $v_\downarrow$ respectively, as
\begin{equation}
\tau_p = \rho_p \phi_b \varphi (v_\uparrow - v_\downarrow),
\label{GrainBornStress}
\end{equation}
where $\phi_b$ is the static bed volume fraction. The trajectories of the grains are controlled by their impact and rebound on the bed. In this layer where the wind is at the threshold, the particles taking off from the bed must have just enough kinetic energy to escape from the potential traps between the static grains. Consequently, the velocities of the saltating grains $v_{\uparrow,\downarrow}$ scale with $\sqrt{gd}$ (Eq.~\ref{EscapeVelocity}), and their typical hop-length is $L \propto d$ \cite{DCA2011}. Under these assumptions, one can then deduce the vertical flux in the feedback layer:
\begin{equation}
\varphi \sim \frac{\rho_f u_t^2}{\phi_b \rho_p \sqrt{gd}} \left( \frac{u_*^2}{u_t^2} - 1 \right)
\label{scalingvarphi}
\end{equation}
and the corresponding transport flux $q=\varphi L$:
\begin{equation}
q \sim \frac{\rho_f u_t^2}{\phi_b \rho_p \sqrt{g/d}} \left( \frac{u_*^2}{u_t^2} - 1 \right).
\label{scalingq}
\end{equation}
Both fluxes scale in a quadratic way with respect to the wind shear stress, in good agreement with experimental and numerical data \cite{RIR1996, DCA2011,HVDO2011,DAC2012,VROD2015}.

\textit{Free wind layer} --- The feedback layer is unstable with respect to the acceleration of the grains that would fly above Bagnold's focal point \cite{A2004}. In this upper region, which we refer to below as the `free wind' layer, the grains can reach the unperturbed wind velocity $v \propto u_*$ and trajectories are of length $L \propto u_*^2/g$, so that the contribution of these grains to the saltation flux would make $q$ scale with the cube of the wind shear stress (Bagnold's scaling \cite{B1941}). This becomes relevant at very strong winds only \cite{DCA2011}. This cubic scaling is also observed for saltation on non-erodible beds \cite{HVDO2011}, where no feedback layer is present. Steady saltation requires a balance in the exchange of grains between the feedback layer and free wind layer, associated with this change of velocity scale, which is the limiting process in the transient case. This phenomenon explains the saturation of the surface transport properties over few $L$ in the feedback layer but the overall erosion of the bed in the tunnel at low pressures, due to an unbalanced free wind layer.

\textit{Collisional layer} --- As introduced by Dur\'an et al. \cite{DCA2014}, aeolian transport involves a third region at the interface between the feedback layer and the static bed, referred below as the `collisional' layer. In this layer, the grains are found to behave in a quasi-2D gaseous-like manner with a typical velocity $\varphi_b$, the basal value of the vertical flux density profile $\varphi(z)$. The associated collisional stress is then $\rho_p \varphi_b^2$. Through a balance with the grain-born stress of the feedback layer $\rho_f (u_*^2 - u_t^2)$, one can deduce
\begin{equation}
\varphi_b \sim u_t \sqrt{\frac{\rho_f}{\rho_p}} \sqrt{\frac{u_*^2}{u_t^2} - 1},
\label{scalingvarphib}
\end{equation}
a formulation which is in good agreement with numerical data \cite{DCA2014}.

%_______________________
\subsection*{Impact ripples}
Following Dur\'an et al. \cite{DCA2014}, the impact ripples emerge at a wavelength $\lambda \propto q/\varphi_b$, which is a length scale associated with aeolian steady transport. Combining Eqs. (\ref{scalingq}) and (\ref{scalingvarphib}), one obtains:
\begin{equation}
\lambda \sim \frac{u_t}{\sqrt{g/d}} \sqrt{\frac{\rho_f}{\rho_p}} \sqrt{\frac{u_*^2}{u_t^2} - 1}.
\label{ScalingRippleslambda}
\end{equation}
Similarly, these authors have shown that the grain velocity scale in the interfacial collisional layer is also representative of the ripple propagation speed $c \sim \varphi_b$ (Eq. \ref{scalingvarphib}). Both $\lambda$ and $c$ thus scale quasi-linearly with wind velocity, in agreement with experimental data \cite{ACP2006}. Adjustment of these scaling laws on our data (Supp Fig.~\ref{fig:RippleScaling}) yields multiplicative factors of $3.4 \times 10^{-3}$ for $c$ and $1.1 \times 10^3$ for $\lambda$. For a fixed typical value of $u_*/u_t \simeq 1.5$, our scalings yield:
\begin{equation}
c \sim u_t \sqrt{\frac{\rho_f}{\rho_p}} = \sqrt{\frac{\tau_f}{\rho_p}},
\label{ScalingRipplescPressure}
\end{equation}
which stays approximately constant with varying pressure (i.e. $\rho_f$) as $\tau_f$ does not vary much upon decreasing the atmospheric pressure. Similarly, the ripple wavelength follows:
\begin{equation}
\lambda \sim \frac{u_t}{\sqrt{g/d}} \sqrt{\frac{\rho_f}{\rho_p}} = \frac{1}{\sqrt{g/d}} \sqrt{\frac{\tau_f}{\rho_p}},
\label{ScalingRippleslambdaPressure}
\end{equation}
which also stays approximately constant for the same reason.

%_______________________
\subsection*{Feedback of sediment transport and impact ripples on the shear velocity}
\begin{figure}[p]
\begin{center}
\includegraphics[width=\linewidth]{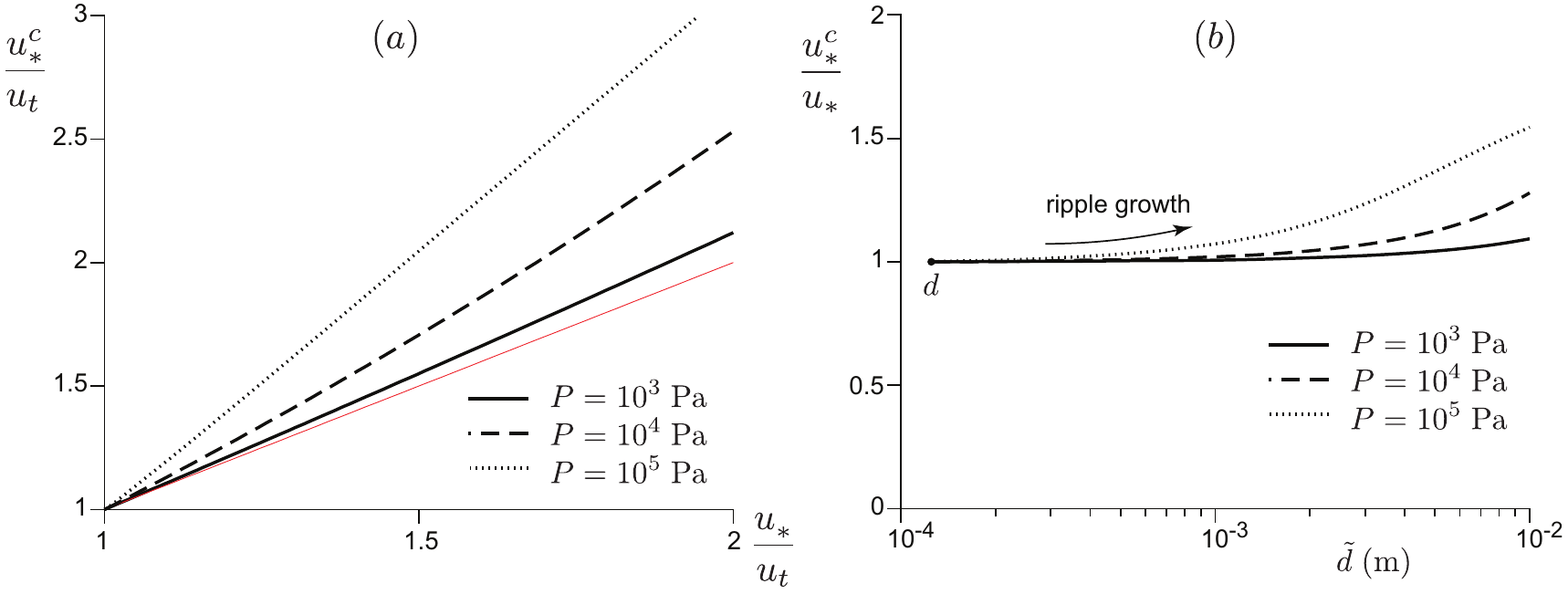}
\end{center}
\caption{(a) Computation of the shear velocity ratio when corrected to account for the presence of sediment transport $u_*^c/u_t$, as a function of the reference value $u_*/u_t$ corresponding to the flat bed composed of fixed grains of diameter $d=125~\mu$m. The line $u_*^c=u_*$ is shown in red. (b) Computation of the factor by which the shear velocity gradually increases due to ripple-induced bed roughening, as parametrized by an equivalent diameter $\tilde d$. In both panels, the different lines correspond to computations for different pressures $P$ (values in legends).}
\label{fig:Roughneffect}
\end{figure}

\begin{figure}[p]
\begin{center}
\includegraphics[width=0.5\linewidth]{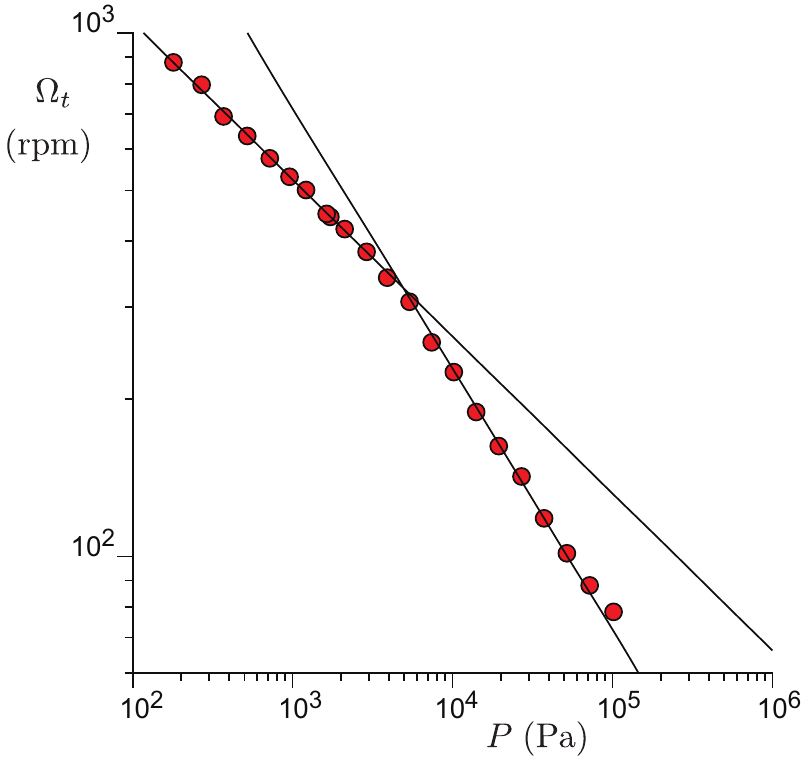}
\end{center}
\caption{Raw data of the threshold fan frequency $\Omega_t$ as a function of pressure $P$. The solid lines are best fits by power laws, showing an exponent $-0.3$ at low pressure and $-0.5$ at high pressure.}
\label{fig:RawData}
\end{figure}

\begin{figure}[p]
\begin{center}
\includegraphics[width=0.9\linewidth]{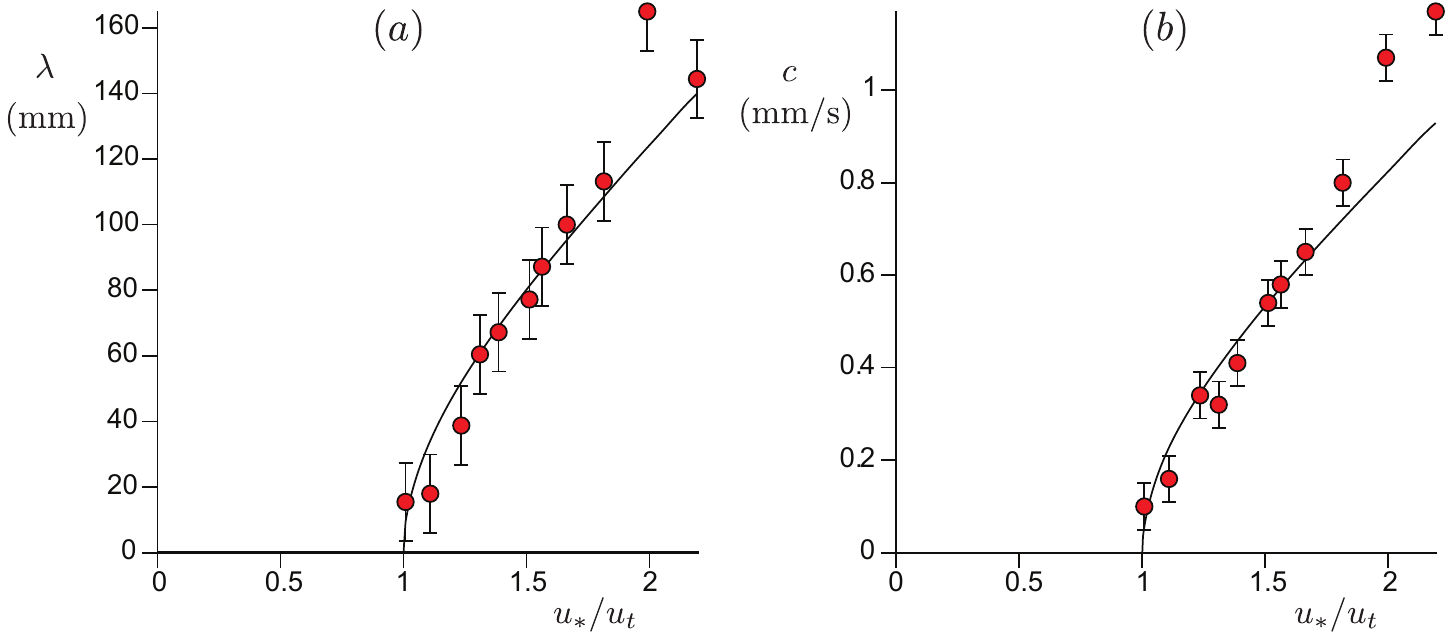}
\end{center}
\caption{Wavelength (a) and propagation speed (b) of impact ripples at ambient pressure (adapted from \cite{ACP2006}). The best fit by Eq. (4-5) of the main text is superimposed, allowing one to extract the multiplicative factors.}
\label{fig:RippleScaling}
\end{figure}

\begin{figure}[p]
\begin{center}
\includegraphics{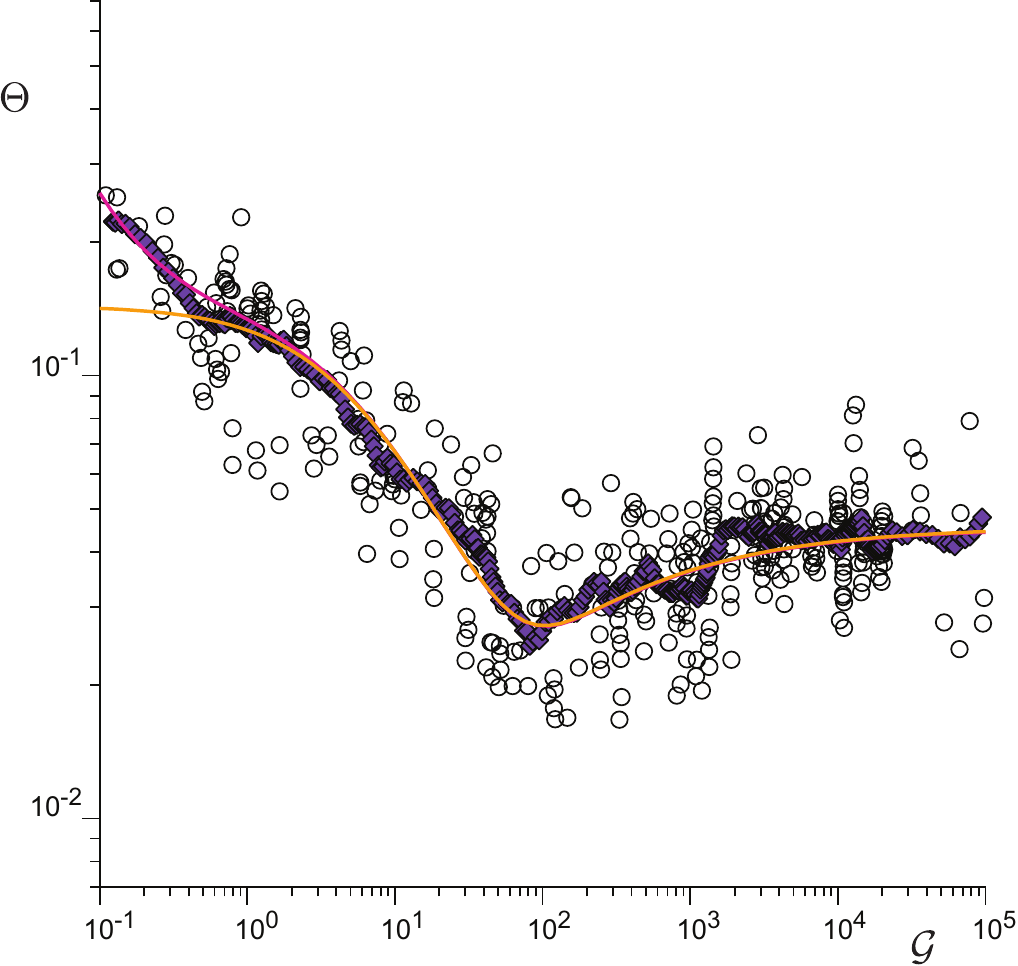}
\end{center}
\caption{Transport threshold under water in the dimensionless plane Shields $\Theta$ \emph{vs} Galileo $\mathcal{G}$ numbers. Circles: data from the literature gathered by \cite{GMM2010}. Diamonds: average over data in a moving window. Solid lines show the model with (violet) and without (orange) cohesion.} 
\label{fig:Water}
\end{figure}

\begin{figure}[p]
\begin{center}
\includegraphics{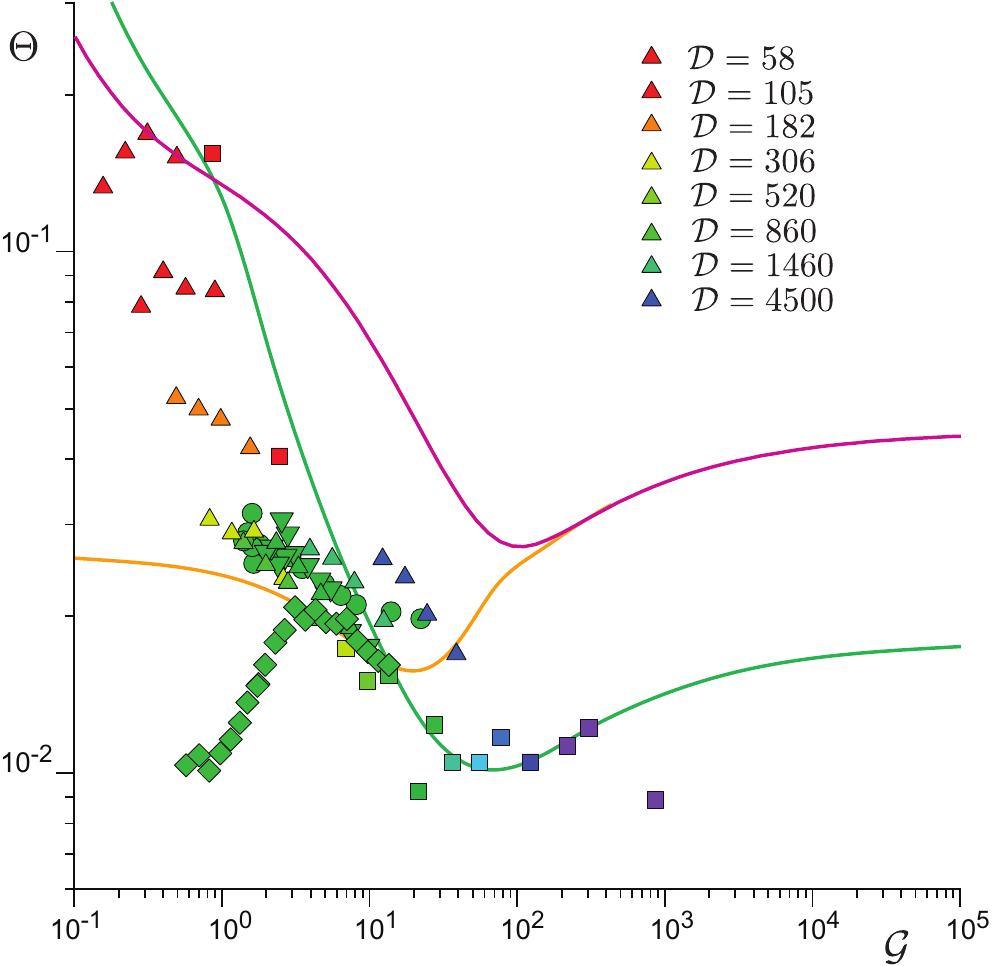}
\end{center}
\caption{Transport threshold in the dimensionless plane Shields $\Theta$ \emph{vs} Galileo $\mathcal{G}$ numbers plotted with all data from \cite{GLWIP1980}. The data  have been obtained for $d=212\;{\mu m}$ in air (circle) and CO$_2$ (down triangle) varying tunnel pressure, and for variable grain sizes $d$ (from top to bottom, $35$, $52$, $75$, $106$, $151$, $212$, $301$ and $641$~$\mu$m) at four different pressures ($5$, $10$, $20$ and $50$~hPa, up triangles). Our data: green diamonds. Saltation threshold data at ambient pressure from the literature \cite{C1945,RIR1996}: squares. Solid lines: same as in Fig. 4. Note that, in contrast with Fig. 4, the symbols' color codes here for the gravitational Stokes number $\mathcal{D}= \frac{1}{\eta} \sqrt{\rho_p (\rho_p-\rho_f)gd^3}=\sqrt{\rho_p/\rho_f} \mathcal{G}$ such that points obtained for the same grain diameter $d$ at different pressures share a same color.}
\label{fig:Greeley}
\end{figure}

\begin{figure}[p]
\begin{center}
\includegraphics{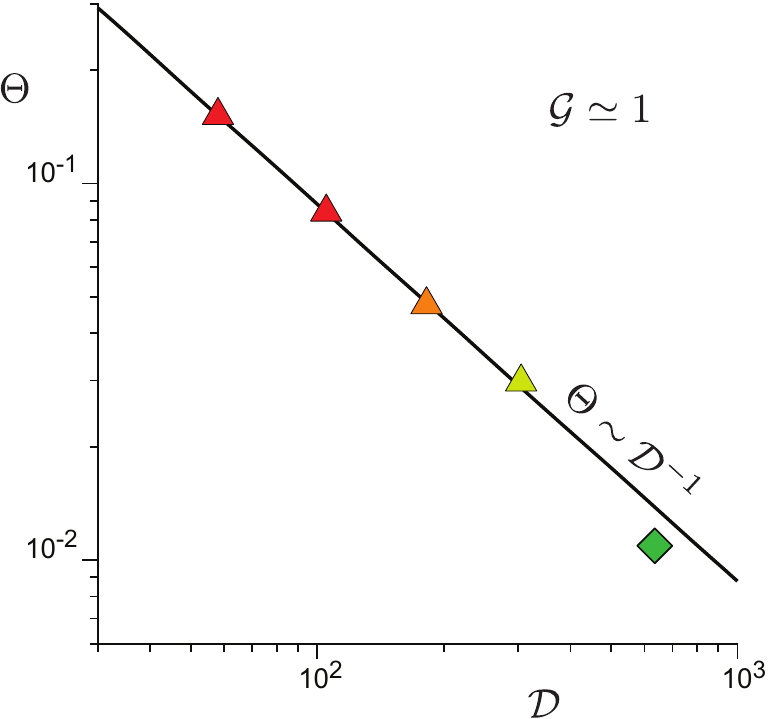}
\end{center}
\caption{Transport threshold in the dimensionless plane relalting the Shields number $\Theta$ to the gravitational Stokes number $\mathcal{D}$ at constant Galileo number around $\mathcal{G} \simeq 1$. As in Supp. Fig. \ref{fig:Greeley}, data from \cite{GLWIP1980} are represented with triangles, and our data with diamonds. The best fit by a power-law gives a phenomenological relation of the form $\Theta \mathcal{D} \simeq 8.8$.}
\label{fig:Greeley_Geq1}
\end{figure}

\begin{figure}[p]
\begin{center}
\includegraphics{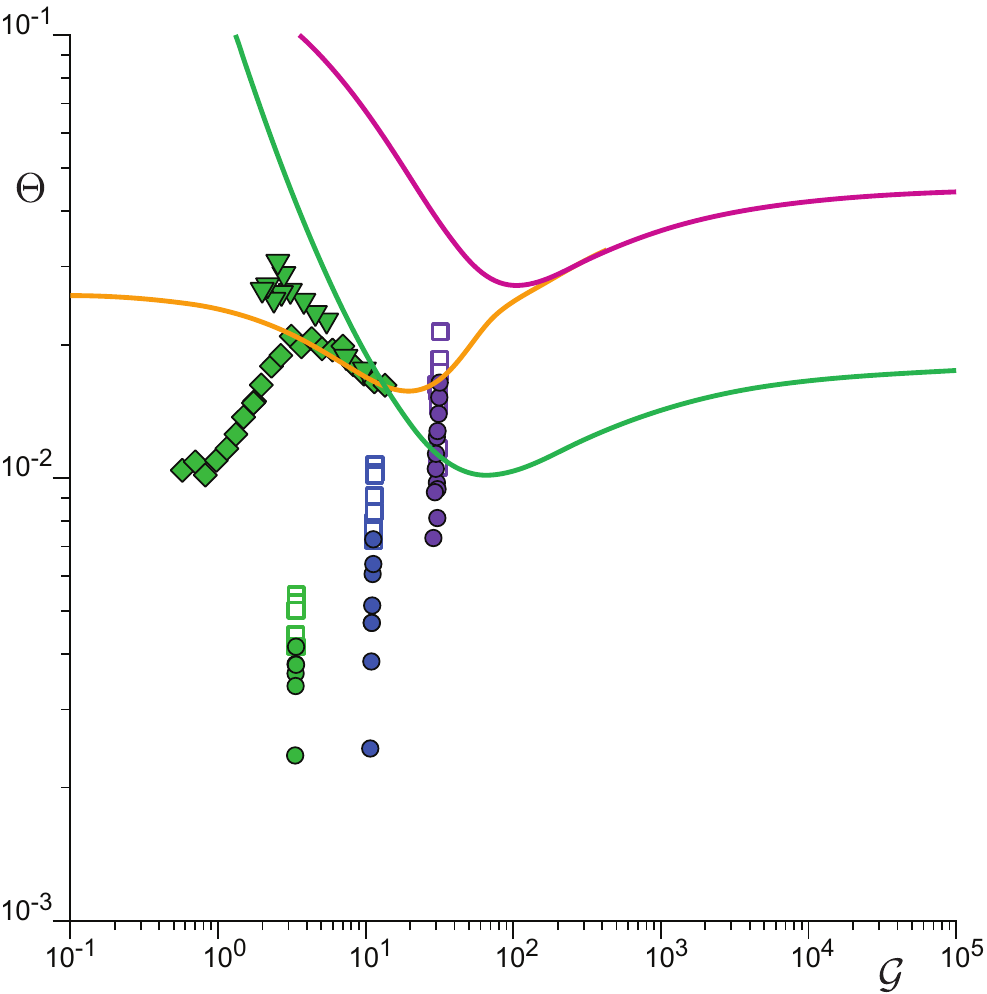}
\end{center}
\caption{Data from \cite{SSE2020} plotted in the dimensionless plane Shields $\Theta$ \emph{vs} Galileo $\mathcal{G}$ numbers with two symbols: their `fluid' (circle) and `general' (square) thresholds for three grain sizes ($310$, $730$ and $1310$~$\mu$m). Note the very large data scatter, associated with qualitative definitions of thresholds: visual observations `as the bed transitioned from intermittent, sporadic motion (fluid threshold) to continuous transport (general threshold)'. For reference: as in previous similar figures, data from \cite{GLWIP1980} are represented with triangles and our data with diamonds. Solid lines and color code: as in previous Supp. Fig.~\ref{fig:Greeley}}
\label{fig:Swann}
\end{figure}

The value of $u_*$ reported in this article always refers to a shear velocity deduced from wind speed profiles obtained for a given fan rotation frequency $\Omega$ above a flat bed made of glued grains without any sediment transport. This is rigorous for the measurements of the sediment transport threshold, performed on a levelled granular surface bed in conditions of residual transport. However, at larger flow speed, the negative feedback of sediment transport on the fluid velocity induces an enhanced bed roughness seen by the flow. Similarly, when impact ripples emerge and develop, their growing amplitude generates a larger effective bed roughness. For a fixed $\Omega$, these two effects increase the shear velocity. In this section, we discuss how to estimate the actual shear velocity $u_*^c$, in comparison to the reference value $u_*$. The results are displayed in Supp. Fig.~\ref{fig:Roughneffect}.

To solve the hydrodynamic model combining Eqs. (\ref{ShearStress},\ref{ellcombo},\ref{ShearStressInTunnel}), we assume that, at a given $\Omega$, the air flow rate and thus the velocity in the central part of the tunnel as well as the pressure gradient along the tunnel remain constant in first approximation. We therefore consider that the vertical gradient of the fluid shear stress $\simeq \rho_f u_*^2/\delta$ is also approximately constant. The model can be solved under these assumptions in the following two cases.

To account for the effect of the feedback of saltation on the wind speed, we simplify the approach of Ungar \& Haff \cite{UH1987} by changing $u_*$ to its threshold value $u_t$ in  (\ref{ShearStress}) in the feedback layer.
The Bagnold's focal point altitude is at the top of the grain trajectory deduced from the dynamic threshold calculation. We plot in Supp. Fig.~\ref{fig:Roughneffect}a the corresponding ratio $u_*^c/u_t$ as a function of the reference value $u_*/u_t$. At low pressure, the curve is close to the diagonal, meaning that the correction is small. This is due to the fact that the transport layer is imbedded into the viscous layer, and has therefore a minor effect on the outer flow. As the pressure $P$ increases, however, the correction is more and more significant. As a consequence, in the experimental runs that we have conducted for the study of ripple emergence with a fan rotation frequency corresponding to $u_*/u_t = 1.5$, the actual value of the shear velocity is likely to be a bit larger, when working close to the ambient pressure.

To model the effect of an increasing aerodynamic roughness during ripple development, we simply replace the actual grain diameter $d$ in Eq. \ref{ellcombo} by an equivalent value $\tilde d$ associated with the ripple amplitude. We plot in Supp. Fig.~\ref{fig:Roughneffect}b the corresponding increase of the ratio $u_*^c/u_*$ as a function of $\tilde d$. This graph shows that the effect of the ripple-induced bed roughness on this ratio decreases with pressure, as a consequence of the fact that the viscous sublayer becomes thicker at smaller $P$. In the experiments, the wavelength and propagation speed of the ripples are measured as soon as they appear, when their amplitude is typically less than a millimeter. We can then neglect that correction to the shear velocity at this stage.

As a conclusion, these roughening effects are negligible for the determination of the sediment transport threshold, which is the main goal of the present paper. They are not so important for the results that we obtained on the characteristics of emerging ripples, but could modify the wavelength and velocity at the higher pressures -- the values of $\lambda$ and $c$ are probably slightly overestimated, as corresponding to effectively larger $u_*/u_t$ than assumed at large $P$. Further experiments using particle image velocimetry are needed to record precise wind profiles above mobile beds.

\begin{table}[p]
\begin{center}
\begin{tabular}{|l|l|l|}
\hline
Grains				& Quartz grains in this study			& Walnut shells				\\
\hline
Bulk density (kg/m$^3$)	&  $\rho_p = 2640 \pm 30$	&  $\rho_p = 1330 \pm 30$	\\
\hline
\end{tabular}
\end{center}
\caption{Bulk density of the quartz grains used in this study, and of the walnut shells used in the NASA Martian wind tunnel.}
\label{Tab:BulkDensityGrains}
\end{table}

\begin{table}[p]
\begin{center}
\begin{tabular}{|c|c|c|c|c|c|}
\hline
Pressure sensor	& Ch1		& Ch2		& Ch4		& Ch5		& Ch7		\\
\hline
G (Pa/count)		& $0.00987$	& $0.00987$	& $0.00984$	& $0.00980$	& $0.00989$	\\
\hline
Standard error ($\times 10^{-4}$)	& $3.50$	& $3.51$	& $3.48$	& $3.46$		& $3.52$\\
\hline
\end{tabular}
\end{center}
\caption{The average sensitivity (G) for five $2.5$~hPa pressure sensors. The values are based values calculated at six pressure values in the range from $3.4$ to $972$~hPa.}
\label{Tab:CalibDiffPressureData}
\end{table}

\begin{table}[p]
\begin{center}
\renewcommand{\arraystretch}{1.2}
\begin{tabular}{| L{0.1cm} L{1.5cm} | L{3.5cm} L{3.3cm} L{1.6cm} L{2.1cm} |}
\hline
\multicolumn{2}{|l|}{\textbf{In wind tunnel}}	& \multicolumn{4}{l|}{Data at ambient pressure and temperature ($P=1$~bar, $T=293$~K)} \\
\cline{3-6}
	& gravity		& $g=9.81$~m/s$^2$ 		&							& 	& \\
	& grains		& $\rho_p=2636$~kg/m$^3$	& $d=125$~$\mu$m				&	& \\
	& air			& $\rho_f=1.204$~kg/m$^3$	& $\eta=1.825 \times 10^{-5}$~Pa s	& $\mathcal{G}=13.5$	& \\
	& CO$_2$		& $\rho_f=1.84$~kg/m$^3$	& $\eta=1.5 \times 10^{-5}$~Pa s	& 	& \\
\hline
\hline
\multicolumn{2}{|l|}{\textbf{On Mars}}			& \multicolumn{4}{l|}{Data representative of Gale crater (b)} \\
\cline{3-6}
	& gravity		& $g=3.7$~m/s$^2$ 					&				& 			& \\
	& grains		& $\rho_p=3000$~kg/m$^3$ \ (a)		&				&			& \\
	& CO$_2$		& $\rho_f=1.89 \times 10^{-2}$~kg/m$^3$	& $P=8.5$~hPa	& $T=243$~K	& \\
\hline
\hline
\multicolumn{2}{|l|}{\textbf{In wind tunnel}}	& \multicolumn{4}{l|}{Data at the Martian density ratio $\rho_p/\rho_f = 1.59 \times 10^5$, but $T=293$~K} \\
\cline{3-6}
	& air 			& $\rho_f=1.66 \times 10^{-2}$~kg/m$^3$	& $P=13.8$~hPa	& $\mathcal{G}=1.58$	& $d_{\rm eff}=140$~$\mu$m \\
	& CO$_2$ 	& $\rho_f=1.66 \times 10^{-2}$~kg/m$^3$	& $P=9.0$~hPa	& $\mathcal{G}=1.93$	& $d_{\rm eff}=160$~$\mu$m \\
\hline
\end{tabular}
\end{center}
\caption{Summary of main data values for fluid and grain mass densities ($\rho_f$, $\rho_p$), fluid dynamic viscosity ($\eta$), grain size ($d$), gravity ($g$) and pressure ($P$). Notes: (a) Estimate of Martian grain composition \cite{Eetal2017}. (b) Martian in-situ measurements by NASA's rover Curiosity \cite{PressureGaleCrater}. The Galileo number $\mathcal{G}$ is here computed for the quartz grains in the wind tunnel, either in the ambient conditions, or at the Martian-like value of the atmospheric density (the two values only differ by the slightly different viscosities of air and CO$_2$). In the Martian context, these $\mathcal{G}$-values correspond to grains of size $d_{\rm eff}=\left[ (\mathcal{G} \eta)^2 / (\rho_f \rho_p g) \right]^{1/3}$.}
\label{AmbientMarsDataTable}
\end{table}

\newpage

%____________________________________________________________________________

\end{document}